\documentclass[lettersize,journal]{IEEEtran}
\usepackage{amsmath,amsfonts}
\usepackage{algorithmic}
\usepackage{algorithm}
\usepackage{array}
\usepackage[caption=false,font=normalsize,labelfont=sf,textfont=sf]{subfig}
\usepackage{textcomp}
\usepackage{stfloats}
\usepackage{url}
\usepackage{verbatim}
\usepackage{color,soul}
\usepackage{graphicx}
\usepackage{hyperref}
\usepackage{cite}
\usepackage[nolist]{acronym}
\usepackage{mathtools}
\usepackage{amssymb}
\usepackage{stfloats}
\usepackage{makecell}
\usepackage{nccmath}
\acrodefplural{HPA}[HPAs]{high power amplifiers}

\begin{acronym}
	\acro{AN}{artificial noise}
	\acro{SI}{self interference}
    \acro{MIMO}{multiple-input, multiple-output}
    \acro{SISO}{single-input, single-output}
    \acro{AWGN}{additive white Gaussian noise}
    \acro{KPI}{key performance indicator}
      \acro{KPIs}{key performance indicators}  
    \acro{D2D}{device-to-device}
    \acro{ISAC}{integrated sensing and communications}
    \acro{DFRC}{dual-functional radar and communication}
    \acro{AoA}{angle of arrival}
    \acro{AoAs}{angles of arrival}
    \acro{MU-MIMO}{multi-user, multiple-input, multiple-output}
    \acro{ToA}{time of arrival}
    \acro{PCA}{principle component analysis}
    \acro{EVM}{error vector magnitude}
    \acro{CPI}{coherent processing interval}
    \acro{eMBB}{enhanced mobile broadband}
    \acro{URLLC}{ultra-reliable low latency comm	unications}
    \acro{mMTC}{massive machine type communications}
	\acro{QCQP}{quadratic constrained quadratic programming}
	\acro{BnB}{branch and bound}
	\acro{SER}{symbol error rate}
	\acro{LFM}{linear frequency modulation}
	\acro{BS}{base station}
	\acro{SLNR}{signal-to-leakage-plus-noise ratio}
	\acro{UE}{user equipment}
	\acro{DL}{deep learning}
	\acro{CS}{compressed sensing}
	\acro{PR}{passive radar}
	\acro{LMS}{least mean squares}
	\acro{NLMS}{normalized least mean squares}
	\acro{ZF}{zero forcing}
	\acro{RIS}{reconfigurable intelligent surface}
	\acro{RISs}{reconfigurable intelligent surfaces}
	\acro{IRS}{intelligent reflecting surface}
	\acro{IRSs}{intelligent reflecting surfaces}
	\acro{NOMA}{non-orthogonal multiple access}
	\acro{TWRN}{two-way relay networks}
	\acro{LISA}{large intelligent surface/antennas}
	\acro{UAV}{unmanned aerial vehicle}
	\acro{LISs}{large intelligent surfaces}
	\acro{OFDM}{orthogonal frequency-division multiplexing}
	\acro{EM}{electromagnetic}
	\acro{ISMR}{integrated sidelobe to mainlobe ratio}
	\acro{MSE}{mean squared error}
	\acro{SNR}{signal-to-noise ratio}
	\acro{IoT}{internet of things}
	\acro{SRP}{successful recovery probability}
	\acro{CDF}{cumulative distribution function}
	\acro{ULA}{uniform linear array}
	\acro{RSS}{received signal strength}
	\acro{SU}{single user}
	\acro{MU}{multi-user}
	\acro{CSI}{channel state information}
	\acro{OFDM}{orthogonal frequency-division multiplexing}
	\acro{DL}{downlink}
	\acro{i.i.d}{independently and identically distributed}
	\acro{UL}{uplink}
	\acro{PSD}{power spectral density}
	\acro{LoS}{line of sight}
	\acro{NLoS}{non line of sight}
	\acro{DFT}{discrete Fourier transform}
    \acro{HPA}{high power amplifier}
    \acro{IBO}{input power back-off}
    \acro{MIMO}{multiple-input, multiple-output}
    \acro{PAPR}{peak-to-average power ratio}
    \acro{AWGN}{additive white Gaussian noise}
    \acro{KPI}{key performance indicator}
    \acro{FD}{full duplex}
    \acro{KPIs}{key performance indicators}  
    \acro{D2D}{device-to-device}
    \acro{ISAC}{integrated sensing and communications}
    \acro{DFRC}{dual-functional radar and communication}
    \acro{AoA}{angle of arrival}
    \acro{AoD}{angle of departure}
    \acro{ToA}{time of arrival}
    \acro{EVM}{error vector magnitude}
    \acro{CPI}{coherent processing interval}
    \acro{eMBB}{enhanced mobile broadband}
    \acro{URLLC}{ultra-reliable low latency communications}
    \acro{mMTC}{massive machine type communications}
	\acro{QCQP}{quadratic constrained quadratic programming}
	\acro{BnB}{branch and bound}
	\acro{mmWave}{millimeter wave}
	\acro{SER}{symbol error rate}
	\acro{LFM}{linear frequency modulation}
	\acro{ADMM}{alternating direction method of multipliers}
	\acro{CCDF}{complementary cumulative distribution function}
	\acro{MISO}{multiple-input, single-output}
	\acro{CSI}{channel state information}
	\acro{LDPC}{low-density parity-check}
	\acro{BCC}{binary convolutional coding}
	\acro{RCS}{radar cross-section}
	\acro{IEEE}{Institute of Electrical and Electronics Engineers}
	\acro{ULA}{uniform linear antenna}
	\acro{B5G}{beyond 5G}
	\acro{MOOP}{multi-objective optimization problem}
	\acro{ML}{maximum likelihood}
	\acro{FML}{fused maximum likelihood}
	\acro{RF}{radio frequency}
	\acro{AM/AM}{amplitude modulation/amplitude modulation}
	\acro{AM/PM}{amplitude modulation/phase modulation}
	\acro{BS}{base station}
	\acro{SCA}{successive convex approximation}
	\acro{MM}{majorization-minimization}
	\acro{LNCA}{$\ell$-norm cyclic algorithm}
	\acro{OCDM}{orthogonal chirp-division multiplexing}
	\acro{TR}{tone reservation}
	\acro{COCS}{consecutive ordered cyclic shifts}
	\acro{LS}{least squares}
	\acro{CVE}{coefficient of variation of envelopes}
	\acro{BSUM}{block successive upper-bound minimization}
	\acro{ICE}{iterative convex enhancement}
	\acro{SOA}{sequential optimization algorithm}
	\acro{BCCD}{block cyclic coordinate descent}
	\acro{SINR}{signal to interference plus noise ratio}
	\acro{MICF}{modified iterative clipping and filtering}
	\acro{PSL}{peak side-lobe level}
	\acro{SDR}{semi-definite relaxation}
	\acro{KL}{Kullback-Leibler}
	\acro{ADSRP}{alternating direction sequential relaxation programming}
	\acro{MUSIC}{MUltiple SIgnal Classification}
	\acro{EVD}{eigenvalue decomposition}
	\acro{SVD}{singular value decomposition}
	\acro{LDGM}{low-density generator matrix}
	\acro{SAC}{sensing and communication}
	\acro{EL}{edge learning}
	\acro{FL}{federated learning}
	\acro{DoF}{degrees-of-freedom}
\end{acronym}

\DeclareMathOperator{\ISMR}{ISMR}
\DeclareMathOperator{\trace}{tr}

\DeclareMathOperator{\SINR}{SINR}
\DeclareMathOperator{\SR}{SR}

\DeclareMathOperator{\dB}{dB}
\DeclareMathOperator{\opt}{opt}

\DeclareMathOperator{\eigvec}{eigvec}
\DeclareMathOperator{\rank}{rank}
\DeclareMathOperator{\EVD}{EVD}
\DeclareMathOperator{\dBm}{dBm}
\DeclareMathOperator{\dBi}{dBi}
\DeclareMathOperator{\meters}{m}
\DeclareMathOperator{\GHz}{GHz}
\DeclareMathOperator{\dBmpHz}{dBm/Hz}
\DeclareMathOperator{\bpspHz}{bits/sec/Hz}
\DeclareMathOperator{\feasibilityrate}{feasibility \ rate}

\DeclareMathOperator*{\argmax}{arg\,max}

\newcommand*\HSI{\pmb{H}_{\tt{SI}}}

\usepackage{xcolor,cite,etoolbox}
\makeatletter 
\pretocmd\@bibitem{\color{black}\csname keycolor#1\endcsname}{}{\fail}
\newcommand\citecolor[1]{\@namedef{keycolor#1}{\color{blue}}}
\makeatother

\begin{document}

\title{Secure Full Duplex Integrated Sensing and Communications}

\author{Ahmad Bazzi,  Marwa Chafii 
\thanks{
Ahmad Bazzi is with the Engineering Division, New York University (NYU) Abu Dhabi, 129188, UAE
(email: \href{ahmad.bazzi@nyu.edu}{ahmad.bazzi@nyu.edu}).

Marwa Chafii is with Engineering Division, New York University (NYU) Abu Dhabi, 129188, UAE and NYU WIRELESS,
NYU Tandon School of Engineering, Brooklyn, 11201, NY, USA (email: \href{marwa.chafii@nyu.edu}{marwa.chafii@nyu.edu}).}
\thanks{Manuscript received xxx}}

\markboth{to appear in IEEE Transactions on Information Forensics and Security, 2024}%
{Shell \MakeLowercase{\textit{et al.}}: A Sample Article Using IEEEtran.cls for IEEE Journals} 

\IEEEpubid{}

\maketitle
\begin{abstract}
The following paper models a secure full duplex (FD) integrated sensing and communication (ISAC) scenario, where malicious eavesdroppers aim at intercepting the downlink (DL) as well as the uplink (UL) information exchanged between the dual functional radar and communication (DFRC) base station (BS) and a set of communication users. The DFRC BS, on the other hand, aims at illuminating radar beams at the eavesdroppers in order to sense their physical parameters, while maintaining high UL/DL secrecy rates. 
Based on the proposed model, we formulate a power efficient secure ISAC optimization framework design, which is intended to guarantee both UL and DL secrecy rates requirements, while illuminating radar beams towards eavesdroppers. The framework exploits artificial noise (AN) generation at the DFRC BS, along with UL/DL beamforming design and UL power allocation. We propose a beamforming design solution to the secure ISAC optimization problem. Finally, we corroborate our findings via simulation results and demonstrate the feasibility, as well as the superiority of the proposed algorithm, under different situations. We also reveal insightful trade-offs achieved by our approach.
\end{abstract}

\begin{IEEEkeywords}
physical layer security, integrated sensing and communications (ISAC), dual-functional radar-communication (DFRC), beamforming design, secrecy rate, artificial noise
\end{IEEEkeywords}

\section{Introduction}
\label{sec:introduction}
\IEEEPARstart{R}{\lowercase{esearch}} has begun to put together a theoretical vision of 6G by looking at potential future services and applications, identifying market demands, and highlighting disruptive technologies \cite{10041914}.  
For instance, \ac{mMTC} expects deploying one million low-price, low-power, and short-range devices, while \ac{URLLC}  targets one milli-second latency to serve mission-critical applications such as autonomous driving and remote robotic surgery.  
Thanks to 6G, the physical and digital worlds will be cohesively and intimately linked,  eventually permeating every aspect of our lives. For that, every network provider that attempts to be trustworthy must address issues including not only high data rates and reliability but also security and privacy. 
According to \cite{dukes2015committee}, security is a state that arises from the creation and maintenance of protective measures that allow an organization to carry out its purpose or fulfill its crucial duties despite the dangers presented by threats to its usage of systems. 
Privacy, on the other hand, is the freedom from intrusion into a person's private life or affairs when that intrusion is caused by the unauthorized or unlawful collection and use of data about that person \cite{garfinkel2016draft}.  
Today, 5G security is a stand-alone architecture, but 6G will shift from a security-only focus to a more comprehensive view of trustworthiness, so it is crucial to integrate security capabilities with \ac{SAC}.

A vital component to enable future 6G systems is \ac{ISAC}, where a common signal is used for \ac{SAC}. 
To further highlight potential jeopardies of unsecure systems, the \ac{CSI} contains enough details revealing critical information on a personal level from both \ac{SAC} perspectives. 
For example, acquiring \ac{CSI} reveals user keystroke information to malicious eavesdroppers \cite{li2016csi}, as well as the means to decode communicating data \cite{9762838}. 
In particular, it is shown in \cite{8613849} that a strong correlation between keystroke gestures and \ac{CSI} fluctuations actually exists. In turn, broadcasting \ac{CSI} information can invade privacy and create many security concerns within wireless networks. 
To address this problem, efforts have been made to secure wirelessly transmitted information. An instance is \cite{jiao2021openwifi}, where authors highlight the importance of privacy and the need to treat sensing parameters, in a private manner. 
For this, a \ac{CSI} fuzzer was proposed so that only the intended receiver can perform \ac{CSI} estimation, thanks to an artificial channel response added to each \ac{OFDM} transmit symbol. 
As a result, the receiver would estimate the overall channel impulse response, which is a cascaded version of the physical channel and the artificial one. Given the knowledge of the latter, the receiver can estimate the former. 
Moreover, wireless transmissions are also vulnerable to security-type attacks, for example, spoofing, due to its broadcast nature. 
Cryptographic approaches do offer a certain level of security in the network, however, the rapid developments made in computing power devices show that even the most mathematically complex secret key-based techniques can be broken, especially when quantum computing becomes a reality \cite{8509094}. 
Therefore, an extra level of security should rely on physical layer design, without the need for secret-key exchange, to maintain secure transmissions especially when both \ac{SAC} information are revealed to malicious users.
\vspace{-0.4cm}
\subsection{Existing work}
\label{sec:existing-work}
To begin with, \cite{4543070} introduces artificial noise injection to enhance the secrecy rate of the system. In precise, artificial noise is added to the signal of interest to enrich the precoding design with additional degrees of freedom to enhance the channel capacity towards legitimate users, while deteriorating it towards malicious ones. For example, \cite{wang2020intelligent} leverage artificial noise injection for a \ac{SU} case, where the \ac{BS}, termed Alice, communicates with the intended communication user, termed Bob, in the presence of a malicious eavesdropper, termed Eve. Although the work aims at maximizing the secrecy rate of the system, the problem is cast as a classical beamforming maximization problem because Eve's rate is treated as a nuisance term throughout the optimization process, due to no prior knowledge of Eve's \ac{CSI}. Furthermore, \cite{6101597} use artificial noise injection, also referred to as masked beamforming therein, for \ac{MU-MIMO} \ac{DL} systems in the presence of only one eavesdropper and outdated \ac{CSI}, whereas \cite{6992154} assumes perfect \ac{CSI}. The work in \cite{6646279} studies hybrid cooperative beamforming strategies tailored for two-way relay networks, with the help of artificial noise injection. It has been found that the scheme not only provides extra array gain, but also blocks the jamming signal from interfering with legitimate users. In addition, the work in \cite{9827562} proposes a \ac{SLNR} design for secure and multi-beam beamforming under a transmit power budget constraint for \ac{DL} satellite communications.
In addition to artificial noise, randomized beamforming was introduced in \cite{6496989}, where all, except the last, beamforming weights are randomized and follow realizations of the Gaussian distribution. The last element allows the legitimate user to perform secure communications, while making it impossible for malicious users to perform channel estimation. 
Researchers have also considered improving security performances through \ac{RIS} technology \cite{9676652, 9383283, 10013699, 10061167,9815181}. For example, a Stackelberg game has been formulated in \cite{9676652} to address energy harvesting concerns for a secure wireless powered communication network, with the aid of \ac{RIS} technology. Also, \cite{9383283} utilizes an \ac{RIS} to improve the secrecy rate of a two-way network, and \cite{10013699} makes use of \ac{RIS} for \ac{NOMA}.
Besides beamforming on a physical layer level to maintain secrecy, efforts have also been made to maintain physical layer security through channel coding. For example, \cite{8252922} investigated the application of \ac{LDGM}, a special type of \ac{LDPC}, on Gaussian wiretap channels. Another instance is \cite{5740591}, where an \ac{LDPC} variant is proposed with finite block lengths that can be combined with cryptographic codes to enhance communication security. Polar codes were also studied in the context of physical layer security \cite{7217814,6891165}.
The aforementioned work exclusively centers on securing communication signals conveying communication information, addressing the paramount concern for security at the physical layer due to the intrinsic broadcast characteristic of wireless signals.
This concern has multiplied following the advent of \ac{ISAC}, creating additional security problems. 
Indeed, integrating sensing capabilities via communication networks unlocks unprecedented new vulnerabilities and security concerns to illegitimate sensing that must be dealt with.
More specifically, \ac{ISAC} signals are designed to contain both sensing, as well as communication information. Therefore, when intending to sense targets, the \ac{ISAC} signal being reused for sensing would be optimized to reach the target with certain radar power characteristics. 
If the \ac{DFRC} or \ac{ISAC} \ac{BS} does not account for possible suspicious activities arising due to malicious targets, then \ac{ISAC} technology can be deemed insecure because of the natural communication information contained within the \ac{ISAC} signal, which can be intercepted by unauthorized targets. Hence, it is crucial to model \ac{ISAC} systems in the presence of potential eavesdroppers so that \ac{ISAC} systems operate at their full effective capacity.
Moreover, it is worth noting that efforts are being made to propose secure \ac{ISAC} techniques. 
For example, \cite{9199556} considers a single eavesdropper case,  leveraging artificial noise to secure the transmitted \ac{ISAC} signal from the \ac{DFRC} \ac{BS} for half-duplex communications, as there are no users participating in the \ac{UL}, and the sensing radar design is done separately via a radar-only problem. 
The work in \cite{9927490,10039359} models a physical layer security \ac{NOMA}-aided \ac{ISAC} system, which exploits \ac{AN}. In particular, a secure precoding approach was designed by maximizing the sum secrecy rate for different users through artificial jamming, and simultaneously utilizing the \ac{NOMA} signals for target detection.Therefore, the \ac{DFRC}/\ac{ISAC} \ac{BS} and users must account for revealing minimal communication information in their transmitted \ac{ISAC} signal in the directions of those possibly malicious targets.
Towards this security concern, the main new technical challenge when designing \ac{ISAC} waveforms in \ac{FD} cases is to carefully optimize the corresponding \ac{ISAC} beamformers and the \ac{ISAC} \ac{AN} statistics in order to attain the three main functionalities in \ac{FD} \ac{ISAC}, namely sensing, communication and secrecy performance, simultaneously. 
In contrast to recent work on \ac{FD} \ac{ISAC}, our work is the first to address security concerns that can arise within \ac{FD} \ac{ISAC}.
In this paper, we are interested in secure \ac{FD} \ac{ISAC} systems.
The \ac{FD} operation refers to a wireless system in which the same time and frequency resources are used by the \ac{DFRC} \ac{BS} transmitting and receiving signals from multiple \ac{DL}/\ac{UL} users, as well as the sensing echo. 
From the perspective of secure \ac{ISAC}, the \ac{DFRC} \ac{BS} optimally transmits a secure \ac{ISAC} signal, containing confidential communication information, and is also designed for secure sensing tasks, whereby sensing beams are formed towards the malicious targets through power focusing, while smartly camouflaging the communication information.
Simultaneously, within the same resources, the \ac{DFRC} \ac{BS} receives the \ac{UL} signal from communication users, superimposed onto the sensing echo.
This \ac{FD} operation facilitates concurrent bi-directional communication and sensing.
Moreover, the \ac{FD} \ac{ISAC} model herein models various sources of interference arising due to the \ac{FD} nature of the problem, such as the \ac{FD} \ac{SI} and the \ac{UL}-to-\ac{DL} interference.
The work in \cite{10158711} has addressed \ac{FD} communication \ac{ISAC} from a power optimization and beamforming perspective; nonetheless, there are many differences that we want to highlight.
From a sensing point of view, we have adopted the \ac{ISMR} radar metric to optimize the sensing performance, which can then enable us to illuminate multiple eavesdroppers, and at the same time reveal minimal communication information towards the mainlobes specified within the \ac{ISMR}. The \ac{ISMR} has a desirable impact not only on target detection, but also on localization performance, where power focusing can be done through the mainlobes pointing towards the eavesdroppers, while keeping low sidelobes to suppress unwanted returns, such as signal-dependent interference and clutter. 
Furthermore, our model captures the presence of possible malicious targets within the scene, which then enables us to define secrecy constraints for the \ac{FD} \ac{ISAC} scenario in both \ac{UL} and \ac{DL}.
For this, we have adopted an \ac{AN} approach, where additional parameters have to be involved in order to optimize the \ac{AN} statistics, hence investing power into \ac{AN} generation. 
Therefore, the beamformers herein, in conjunction with \ac{AN} statistics and \ac{UL} power levels, are designed not only to enhance communication rates, but also to improve \ac{UL} and \ac{DL} secrecy rates, \textit{and concisely illuminate signal power in target directions without revealing the communication information embedded in the transmitted \ac{ISAC} signal.}
In addition, the work in \cite{10159012} mainly focuses on an \ac{FD} \ac{ISAC} setting to model the “echo-miss” problem, which is a case arising from strong residual \ac{SI} that can dominate the radar echo, hence missing the backscattered echo due to target presence. To solve this, the authors in \cite{10159012} proposes an optimization problem to satisfy \ac{UL}/\ac{DL} communication data rates and generate transmit, receive and post-\ac{SI} beamformers. 
It is worth noting that our work also differs in several aspects. 
First, the sensing metric adopted in \cite{10159012} is the radar beampattern power output for one and only one target, without acknowledging clutter components.
Second, and as previously mentioned, we model potential communication leaks onto multiple radar targets, which then allows us to define relevant security metrics such as \ac{UL} and \ac{DL} secrecy rates in our \ac{FD} \ac{ISAC} scheme.

\vspace{-0.5cm}
\subsection{Contributions and Insights}
\label{sec:contributions-insights}
This work focuses on a secure \ac{FD} design, where the \ac{DFRC} \ac{BS} transmitting \ac{ISAC} signals, designed for \ac{SAC}, aims at simultaneously securing the \ac{ISAC} signal in the \ac{DL} and \ac{UL}, in the presence of multiple eavesdroppers. To that purpose, we have summarized our contributions as follows.
\begin{itemize}
	\item \textbf{Multi-user Secure Full Duplex Model}. We model an \ac{FD} scenario that accommodates multiple legitimate \ac{DL} and multiple legitimate \ac{UL} communication users associated with a \ac{DFRC} \ac{BS}. Moreover, we assume multiple malicious eavesdroppers that are capable of intercepting both \ac{DL} and \ac{UL} communication signals exchanged between the users and \ac{DFRC} \ac{BS}. 
	\item \textbf{Secure \ac{ISAC} Optimization Framework}. Due to the \ac{FD} protocol, we leverage artificial noise injection not only to improve the secrecy in the \ac{DL}, but also in the \ac{UL}, hence deteriorating eavesdroppers reception for both \ac{UL}/\ac{DL} signals. As will be shown, the artificial noise aids in creating radar beams toward the eavesdroppers so that the \ac{DFRC} \ac{BS} can physically sense the eavesdroppers. A power-efficient secure \ac{ISAC} optimization problem is designed to achieve all the aforementioned tasks. 
	\item \textbf{Solution via successive convex approximations}. Due to the non-convex nature of the secure \ac{ISAC} optimization problem at hand, we resort to a \ac{SCA} technique, which enables us to exploit the non-convex optimization problem and solve it as a series of convex optimization problems. The resulting algorithm is iterative and converges with a few iterations. 
	\item \textbf{Extensive simulation results}. We present extensive simulation results that demonstrate the superiority, as well as the potential of the proposed design and algorithm with respect to benchmarks, in terms of secrecy rates, and total power consumption of the \ac{FD} system.   
\end{itemize}
Furthermore, we unveil some important insights, i.e.
\begin{itemize}
	\item The proposed algorithm enjoys fast convergence properties, as it requires as little as $20$ iterations to settle at a stable solution.	
	\item We prove $\rank$-one optimality of the \ac{DL} beamforming vectors at each iteration of the proposed \ac{SCA} method. Indeed, this property is highly desired when a $\rank$ relaxation has been imposed at some point of an optimization problem. The advantage here is that there is no longer a necessity for post-processing methods, such as Gaussian randomization, to approximation $\rank-$one solutions. 
	\item For a fixed number of eavesdroppers, the secrecy rate can settle to a stable value regardless of the number of \ac{UL} and \ac{DL} communication users. As compared to state-of-the-art methods, the secrecy rate can be controlled through design parameters making it less sensitive to the number of \ac{UL} and \ac{DL} users. 
	\item We unveil fruitful trade-offs in terms of secure communications versus radar sensing that can be achieved with the proposed method. We show through simulations that decreasing \ac{ISMR} to achieve better radar sensing performance deteriorates the secrecy performance. In addition to secure \ac{ISAC} trade-offs, we also discuss power-secrecy trade-offs.
\end{itemize}

\vspace{-0.5cm}
\subsection{Organization and Notations}
The detailed structure of this paper is given as follows: Section \ref{sec:system-model} presents the full-duplex system model and the key performance indicators adopted in this paper. In Section \ref{sec:opt-probs}, we formulate a suitable optimization framework tailored for the secure \ac{FD} \ac{ISAC} design problem. The proposed algorithm is presented in Section \ref{sec:opt-01}. Our simulation findings are given in Section \ref{sec:simulations}. Finally, we conclude the paper in Section \ref{sec:conclusions}.

\textbf{Notation}: Upper-case and lower-case boldface letters denote matrices and vectors, respectively. $(.)^T$, $(.)^*$, and $(.)^H$ represent the transpose, the conjugate, and the transpose-conjugate operators. The statistical expectation is denoted as $\mathbb{E}\lbrace \rbrace$. For any complex number $z \in \mathbb{C}$, the magnitude is denoted as $\vert z \vert$, its angle is $ \angle z$. The $\ell_2$ norm of a vector $\pmb{x}$ is denoted as $\Vert \pmb{x} \Vert$. The matrix $\pmb{I}_N$ is the identity matrix of size $N \times N$. The zero-vector is $\pmb{0}$. For matrix indexing, the $(i,j)^{th}$ entry of matrix $\pmb{A}$ is denoted by $[\pmb{A}]_{i,j}$ and its $j^{th}$ column is denoted as $[\pmb{A}]_{:,j}$. The operator $\EVD$ stands for eigenvalue decomposition. The projector operator onto the space spanned by the columns of matrix $\pmb{A} \in \mathbb{C}^{N \times M}$ is $\pmb{P}_{\pmb{A}}$ and the corresponding orthogonal projector is $\pmb{P}_{\pmb{A}}^{\perp} \triangleq \pmb{I} - \pmb{A}(\pmb{A}^H \pmb{A})^{-1} \pmb{A}^H$. The $[x]^+$ operator returns the maximum between $x$ and $0$. A positive semi-definite matrix is denoted as $\pmb{A} \succeq \pmb{0}$ and a vector $\pmb{x}$ with all non-negative entries is denoted as $\pmb{x} \succeq \pmb{0}$. The all-ones vector of appropriate dimensions is denoted by $\pmb{1}$.

\vspace{-0.2cm}
\section{System Model}
\label{sec:system-model}
\begin{figure}[!t]
\centering
\includegraphics[width=3.5in]{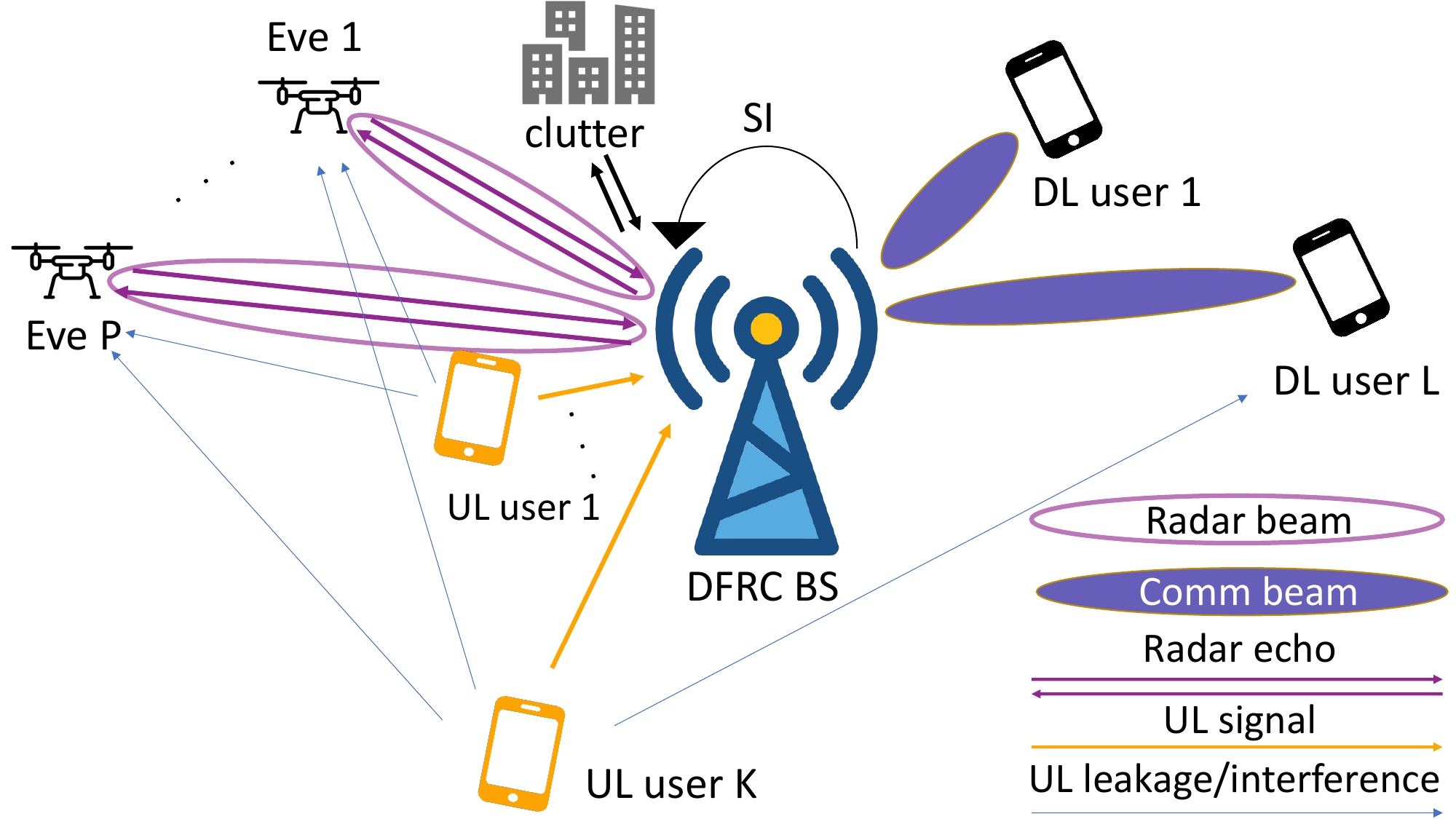}
\caption{A practical \ac{FD} \ac{ISAC} scenario, in the presence of multiple malicious eavesdroppers intending to intrude on \ac{UL} and \ac{DL} communications between legitimate \ac{UL}/\ac{DL} users and a \ac{DFRC} \ac{BS}. Note that only one \ac{UL}-to-\ac{DL} interfering component is depicted, due to presentation sake.}
\label{fig_1}
\end{figure}
\subsection{Full Duplex Radar and Communication Model}
\label{sec:FD-DFRC-model}
Let us assume an \ac{FD} \ac{ISAC} scenario, where a threat of intercepting data by malicious eavesdroppers exists in the scene. To this extent, let us first consider a \ac{DFRC} \ac{BS}, equipped with $N_T$ transmitting antennas and $N_R$ receiving antennas displaced via a mono-static radar setting with no isolation between both arrays. 
In baseband, the \ac{DFRC} \ac{BS} transmits the following $N_T \times 1$ secure \ac{ISAC} signal vector
\useshortskip
\begin{equation}
	\label{eq:Tx-signal-BS}
	\pmb{x} = \pmb{V}\pmb{s} + \pmb{w},
\end{equation}
where $\pmb{V} \in \mathbf{C}^{N_T \times L}$ is the \ac{DFRC} beamforming matrix, where its $\ell^{th}$ column is loaded with the beamforming vector dedicated towards the $\ell^{th}$ communication user. In addition, the total number of communication users in the \ac{DL} is denoted by $L$. Moreover, $\pmb{s}$ is the transmit data symbols, where its $\ell^{th}$ entry is the data symbol intended to the $\ell^{th}$ communication \ac{DL} user.  In what follows, the symbols are assumed to be independent with unit variance, i.e. $\mathbb{E}(\pmb{s}\pmb{s}^H) = \pmb{I}_L$. Furthermore, a superimposed \ac{AN} vector, which is assumed to be independent from the signal $\pmb{s}$, is transmitted via the following statistical characteristics,
\useshortskip
\begin{equation}
	\label{eq:AN}
	\pmb{w} \sim \mathcal{N}(\pmb{0},\pmb{W}),
\end{equation}
where $\pmb{W} \succeq \pmb{0}$ stands for the spatial \ac{AN} covariance matrix, which is not given and, needs to be computed. Also, $\pmb{w}$ and $\pmb{s}$ are assumed to be independent of one another. Note that $\trace(\pmb{W})$ would then reflect the total transmit power invested on \ac{AN}. It is worth noting that the secure \ac{ISAC} signal $\pmb{x}$ is not only intended for the \ac{DL} users, but also is designed to be a radar signal through $\pmb{V}$. Hence, $\pmb{V}$ is designed for \ac{DFRC} beamforming to maintain good communication properties, while preserving some radar properties that will be clarified in the paper. Moreover, the secure \ac{ISAC} signal $\pmb{x}$ is "secure" in a sense that it is \ac{AN}-aided through $\pmb{w}$ as defined in equation \eqref{eq:AN}. It is pertinent to highlight that vector $\pmb{w}$ serves a dual purpose, being employed not only for \ac{AN} purposes but also to augment the \ac{DoF} of the transmitted signal, thereby improving sensing performance.
More precisely, and in what follows, the optimization process involves the participation of the covariance of $\pmb{w}$, i.e. $\pmb{W}$, to achieve a desired \ac{ISMR}, which has positive reverberations on  target detection and localization performance.We note that, as per \ac{ISAC} design, this model does not specifically separate the communication and sensing signals, and so regards them as part of the same signal.Hence, the single-antenna \ac{DL} communication users receive the following signal
\begin{equation}
	\label{eq:DL-signal-1}
		{y}^{\tt{DL}}_\ell
	=
	\sum\nolimits_{k=1}^K
	q_{k,\ell}v_k
	+
	\pmb{h}_{r,\ell}^H
	\pmb{x}
	+
	n_\ell, \quad \ell = 1 \ldots L
\end{equation}
where $\pmb{h}_{r,\ell}$ denotes the \ac{mmWave} channel between the $\ell^{th}$ \ac{DL} communication user and the \ac{DFRC} \ac{BS} modeled as \cite{7914742}
\useshortskip
\begin{equation*}
	\pmb{h}_{r,\ell}
=
\sqrt{\frac{\kappa_\ell}{\kappa_\ell + 1}}
\pmb{h}_{r,\ell}^{\tt{LoS}}
+
\sqrt{\frac{1}{\kappa_\ell + 1}}
\pmb{h}_{r,\ell}^{\tt{NLoS}}
\in 
\mathbb{C}^{N_T \times 1}
,
\end{equation*}
where $\kappa_\ell$ is the Rician $K$-factor of the $\ell^{th}$ user.
Moreover, $\pmb{h}_{r,\ell}^{\tt{LoS}}$ is the \ac{LoS} component between the $\ell^{th}$ user and the \ac{DFRC} \ac{BS}.
In addition, the \ac{NLoS} component consists of $N_{\mathrm{cl}}$ clusters as follows
\useshortskip
\begin{equation*}
	\pmb{h}_{r,\ell}^{\tt{NLoS}}  =
\sqrt{\frac{1}{\sum_{q=1}^{N_{\mathrm{cl}}} N_{q}}} \sum_{q=1}^{N_{\mathrm{cl}}} \sum_{r=1}^{N_{q}} u_{q,r} 
\pmb{a}_{N_T}(\varphi_{q,r}),

\end{equation*}
where $N_q$ is the number of propagation paths within the $q^{th}$ clusters.
In addition, $u_{q,r}$ and $\varphi_{q,r}$ denote the path attenuation and \acp{AoD} of the $r^{th}$ propagation path within the $q^{th}$ cluster.
As the number of clusters grows, the path attenuation coefficients and angles between users and the \ac{DFRC} \ac{BS} become randomly distributed. Consequently, we model the entries of $\pmb{h}_{r,\ell}^{\tt{NLoS}}$ as \ac{i.i.d} complex Gaussian random variables \cite{7914742}.In addition, $q_{k,\ell}$ is the channel coefficient between the $k^{th}$ \ac{UL} communication user and the $\ell^{th}$ \ac{DL} user. The interference caused by \ac{UL} onto \ac{DL} can be detrimental when \ac{DL}/\ac{UL} users are nearby one another.
Moreover, the $k^{th}$ \ac{UL} symbol formed by the $k^{th}$ user and intended towards the \ac{DFRC} \ac{BS} is denoted as $v_k \in \mathbb{C}$. 
The power level of $v_k$ is $\mathbb{E}(\vert v_k \vert^2) = p_k, \forall k$.
Due to clutter presence, a portion of the \ac{DL} signal scatters away from the clutter towards the \ac{DL} communication users. Therefore, the \ac{DL} channels $\pmb{h}_{r,\ell}$ account for clutter.We can re-express \eqref{eq:DL-signal-1} as
\useshortskip
 \begin{equation}
 	\label{eq:DL-signal-2}
		{y}^{\tt{DL}}_\ell
	=
	\pmb{h}_{r,\ell}^H
	\pmb{v}_\ell s_\ell
	+
	\sum\limits_{k=1}^K
	q_{k,\ell}v_k
	+
	\sum\limits_{\substack{\ell^{'} \neq \ell}}^L
	\pmb{h}_{r,\ell}^H
	\pmb{v}_{\ell^{'}} s_{\ell^{'}}
	+
	\pmb{h}_{r,\ell}^H
	\pmb{w}
	+
	n_\ell,
 \end{equation}
where the first term represents the desired part of the received signal, 
the second term reflects the \ac{UL}-to-\ac{DL} interference,
the third term is the multi-user interference, 
the fourth noise is the jamming \ac{AN} signal part and 
$n_\ell$ is zero-mean \ac{AWGN} noise with variance $\sigma_{\ell}^2$, i.e. $n_\ell \sim \mathcal{N}(0,\sigma_{\ell}^2)$.
  
The eavesdroppers are not only intercepting the \ac{UL} signals, but also the \ac{DFRC} \ac{BS} \ac{DL} signals. Thus, the signal at the $p^{th}$ eavesdroppers can be written as
\useshortskip
\begin{equation}
	\label{eq:pth-eve}
	y_p^{\tt{Eve}}
	=
	\sum\nolimits_{k=1}^K
	g_{k,p}v_k
	+
	\alpha_p
	\pmb{a}_{N_T}^H(\theta_p)
	\pmb{x}
	+
	z_p, \quad p = 1 \ldots P
\end{equation} 
where $g_{k,p}$ is the complex channel coefficient between the $k^{th}$ \ac{UL} communication user and the $p^{th}$ eavesdroppers. Moreover, $\alpha_p$ corresponds to the path-loss complex coefficient between the \ac{DFRC} \ac{BS} and the $p^{th}$ eavesdropper and $\pmb{a}(\theta)$ is the steering vector pointing towards angle $\theta$, i.e.
\useshortskip
\begin{align}
	\label{eq:steering-vector}
	\pmb{a}_{N_T}(\theta)
	&=
	\begin{bmatrix}
		a_1(\theta) & \ldots & a_{N_T}(\theta) 
	\end{bmatrix}^T, \\
	\label{eq:steering-vector-2}
	\pmb{a}_{N_R}(\theta)
	&=
	\begin{bmatrix}
		a_1(\theta) & \ldots & a_{N_R}(\theta) 
	\end{bmatrix}^T. 
\end{align}
For example, in a \ac{ULA} setting, we have that $a_n(\theta) = e^{j \frac{2\pi}{\lambda} (n-1) d \sin(\theta)}$, where $\lambda$ is the wavelength. 
The eavesdropper \acp{AoA} are denoted as $\theta_p$.
Moreover, the \ac{AWGN} at the $p^{th}$ eavesdropper is $z_p$, which is modeled as zero-mean and variance $\sigma_p^2$, i.e. $z_p \sim \mathcal{N}(0,\sigma_p^2)$.
We would like to highlight the relationship between the received signal at the $\ell^{th}$ legitimate \ac{DL} user in \eqref{eq:DL-signal-1} (equivalently \eqref{eq:DL-signal-2}) and the eavesdropper in \eqref{eq:pth-eve}. 
We first note that both legitimate and eavesdroppers receive the \ac{UL} signals, as well as the \ac{DFRC} \ac{ISAC} signal $\pmb{x}$.
However, a different channel model is adopted between the \ac{DFRC}-\ac{DL}, i.e. $\pmb{h}_{r,\ell}$, and \ac{DFRC}-eavesdropper, i.e. $\pmb{a}(\theta_p)$.
This is because we regard eavesdroppers as radar targets, where a strong and dominant \ac{LoS} component exists between the \ac{DFRC} and the eavesdroppers (Rician channel with large $K-$factor) for localization and tracking applications.
This model is applicable to scenarios involving \acp{UAV} \cite{9916163}, \cite{10098686}.
More specifically, the \ac{DFRC} aims at inferring additional sensing information about eavesdropping targets in the scene (such as delay and doppler), without exposing the communication information embedded within the transmit \ac{ISAC} signal, $\pmb{x}$.
Meanwhile, a set of $K$ \ac{UL} communication users associated with the \ac{DFRC} \ac{BS} co-exist. It is important to note that while the \ac{DFRC} \ac{BS} transmits a secure signal vector $\pmb{x}$, it concomitantly collects the \ac{UL} signals, along with other components that we shall clarify in this section, as well. Furthermore, let $\pmb{h}_{t,k}$ be the \ac{UL} channel vector between the $k^{th}$ \ac{UL} communication user and the \ac{DFRC} \ac{BS}. For Rayleigh fading, the \ac{UL} channels $\pmb{h}_{t,k}$ also account for the clutter present in the environment, where part of the channel propagates towards the clutter and back to the \ac{DFRC} \ac{BS}.To this end, the signal received by the \ac{DFRC} \ac{BS} due to the single-antenna \ac{UL} users is $\sum\nolimits_{k=1}^K \pmb{h}_{t,k} v_k$.

We summarize the system model in Fig. \ref{fig_1}, which shows a group of \ac{UL} and \ac{DL} legitimate communication users associated with a \ac{DFRC} \ac{BS}. In addition, a group of eavesdroppers listen to \ac{UL} and \ac{DL} signal exchanges. Note that the \ac{DL} signal is an \ac{ISAC} signal with the \ac{DFRC} \ac{BS} highlights beams towards the eavesdroppers in order to acquire physical sensing information about the eavesdroppers. 
Moreover, due to \ac{FD} operation, the simulateneous transmission of signal vector $\pmb{x}$ is \textit{"leaked"} to the receiving unit of the \ac{DFRC} \ac{BS}, in a form of loop interference channel. This phenomenon is referred to as \textit{\ac{SI}} and is one crucial component when targeting \ac{FD} behavior, in particular when the transmitting and receiving units of the \ac{DFRC} \ac{BS} are colocated, without isolation. Indeed, the receiving unit sees an \ac{SI} component of $\sqrt{\beta}\HSI \in \mathbb{C}^{N_R \times N_T}$, where $\beta$ models the residual \ac{SI} power and under the direct-coupling channel model for self-interference, we have \cite{9385108}
\useshortskip
\begin{equation}
	\label{eq:SI}
	[\HSI]_{i,j}
	=
	\exp \Big( j 2 \pi \frac{d_{i,j}}{\lambda} \Big),
\end{equation}
contains the phases introduced between each pair of transmitting and receiving antennas. Note that $d_{i,j}$ is the distance between the $i^{th}$ receive and $j^{th}$ transmit antennas. Furthermore, the \ac{DFRC} \ac{BS} aims at illuminating radar power towards the $P$ eavesdroppers. So, the echo can be modeled as 
\useshortskip
\begin{equation}
	\label{eq:echo}
	\pmb{y}_r = 
	\sum\nolimits_{p=1}^P
	\gamma_p \pmb{a}_{N_R}(\theta_p)\pmb{a}_{N_T}^H(\theta_p)\pmb{x} + \pmb{c}.
\end{equation} 
The complex amplitude due to the $p^{th}$ eavesdropper is $\gamma_p$ and $\pmb{c} \sim \mathcal{N}(\pmb{0},\pmb{R}_c)$ is the clutter including the reflected signal from the \ac{UL}/\ac{DL} communication users \cite{9724174}. 
The clutter covariance matrix $\pmb{R}_c \in \mathbb{C}^{N_R \times N_R}$ capturing all environmental clutter effects is assumed to be constant. 
The eavesdropper \acp{AoA} are assumed to be knowns. 
This \textit{a priori} knowledge can be acquired through a previous estimation stage, whereby the \ac{DFRC} \ac{BS} uses this knowledge to design a secure \ac{FD} \ac{ISAC} signal.
Indeed, the \ac{DFRC} \ac{BS} aims at estimating additional sensing parameters, like the delay for distance estimation and doppler for speed estimation, which is of interest for localization and tracking applications.
In assuming partial knowledge through \acp{AoA}, the \ac{DFRC} \ac{BS} can further optimize its transmit signal so that the received echo is good enough, from a localization accuracy perspective \cite{9724174}, to further estimate/track additional sensing parameters, e.g. delay and doppler.
Other \ac{ISAC} works that utilize \ac{AoA} target knowledge include \cite{9916163} and \cite{9724174}.It is worth noting that, in equation \eqref{eq:echo}, due to the collocated transmit and receiving units, the \ac{AoA} and \ac{AoD} are the same, as vectors $\pmb{a}_{N_R}(\theta_p)$ and $\pmb{a}_{N_T}^T(\theta_p)$ appearing in the echo are parametrized through the same angle. Finally, the received signal at the \ac{DFRC} \ac{BS} is
\vspace{-0.2cm}
\begin{equation}
	\label{eq:BS-signal}
	\pmb{y}^{\tt{DFRC}}
	=
	\sum\nolimits_{k=1}^K \pmb{h}_{t,k} v_k
	+
	\pmb{y}_r
	+
	\sqrt{\beta}\HSI \pmb{x}
	+
	\pmb{n}.
\end{equation}
In summary, the first part in \eqref{eq:BS-signal} is the signal due to the $K$ \ac{UL} communication users. The second part highlights the received echo from the $P$ eavesdroppers, as well as environmental clutter. The third part is the \ac{SI}. Finally, the \ac{AWGN} vector is $\pmb{n} \sim \mathcal{N}(\pmb{0},\sigma_n^2 \pmb{I}_{N_R})$.
	

\vspace{-0.5cm}
\subsection{Key Performance Indicators}
In order to address the secure \ac{ISAC} \ac{FD}, it is vital to define \ac{KPIs} that turn out to be essential for assessing the performance of the overall system, as well as designing a suitable optimization framework.

\subsubsection{Communication Achievable Rates}
First, we define the \ac{SINR} of the $\ell^{th}$ \ac{DL} communication user. Based on equation \eqref{eq:DL-signal-2}, and under the assumption of independent symbols, we can write
\begin{equation}
\label{eq:DL-interference-before-expectation}
\begin{split}
&\SINR_{\ell}^{\tt{DL}}
\\&	= 
	\frac{\mathbb{E} \big[\vert \pmb{h}_{r,\ell}^H
	\pmb{v}_\ell s_\ell \vert^2 \big]} 
	     {\sum\limits_{\substack{k = 1  }}^K \mathbb{E}[\vert q_{k,\ell} v_k \vert^2] + \sum\limits_{\substack{\ell^{'} = 1\\ \ell^{'} \neq \ell}}^L \mathbb{E} \big[ \vert \pmb{h}_{r,\ell}^H
	\pmb{v}_{\ell^{'}} s_{\ell^{'}} \vert^2 \big] + \mathbb{E} \big[\vert  \pmb{h}_{r,\ell}^H
	\pmb{w}  \vert^2 \big] +\sigma_{\ell}^2},
\end{split}
\end{equation}
which can also be expressed as
\begin{equation}
	\label{eq:SINR_DL}
		\boxed{
	\SINR_{\ell}^{\tt{DL}}
	= 
	\frac{ \pmb{h}_{r,\ell}^H\pmb{V}_\ell\pmb{h}_{r,\ell} } 
	     {\sum\limits_{\substack{k = 1  }}^K p_k \vert q_{k,\ell} \vert^2 +\sum\limits_{\substack{\ell^{'} = 1\\ \ell^{'} \neq \ell}}^L \pmb{h}_{r,\ell}^H\pmb{V}_{\ell^{'}}\pmb{h}_{r,\ell} + \pmb{h}_{r,\ell}^H\pmb{W}\pmb{h}_{r,\ell}  +\sigma_{\ell}^2}, 
	     }
\end{equation}
where $\pmb{V}_{\ell} = \pmb{v}_{\ell} \pmb{v}^H_{\ell}$. In this case, and according to Shannon-theory, the achievable rate of transmission of the $\ell^{th}$ \ac{DL} communication user is
\vspace{-0.2cm}
\begin{equation}
\label{eq:rate-equation}
	R_{\ell}^{\tt{DL}} = \log_2( 1 + \SINR_{\ell}^{\tt{DL}} ), \quad \forall \ell = 1 \ldots L
\end{equation}
Meanwhile, we also define an \ac{UL} \ac{SINR} to assess \ac{UL} communication performance as seen at the \ac{DFRC} \ac{BS}. Assuming that the \ac{BS} applies $K$ beamforming vectors $\pmb{u}_1 \ldots \pmb{u}_K$, we can define a receive \ac{SINR} quantifying the performance between the $k^{th}$ \ac{UL} communication user and the \ac{DFRC} \ac{BS} as follows
\begin{equation}
	\label{eq:SINR_UL1_first}
	\begin{split}
	&\SINR_{k}^{\tt{UL}}
	\\ &=
	\frac
	{\mathbb{E}[\vert \pmb{u}_k^H \pmb{h}_{t,k} v_k \vert^2]}
	{\sum\nolimits_{\substack{k' = 1 \\ k' \neq k }}^K \mathbb{E}[\vert \pmb{u}_{k}^H \pmb{h}_{t,k'} v_{k'} \vert^2] +
	\mathbb{E}[\vert  \pmb{u}_k^H \pmb{C}\pmb{x}  \vert^2] + 
	\mathbb{E}[\vert \pmb{u}_k^H \pmb{c}  \vert^2] +
	\sigma_n^2},
\end{split}
\end{equation}
where $\pmb{C} = \sum\nolimits_{p=1}^P \gamma_p \pmb{a}_{N_R}(\theta_p)\pmb{a}_{N_T}^H(\theta_p) + \sqrt{\beta}\HSI $. After applying the expectation operator, we can rewrite \eqref{eq:SINR_UL1_first} as the equation given in equation \eqref{eq:SINR_UL1}.
\useshortskip
 \begin{figure*}[b]
 \begin{equation}
	\label{eq:SINR_UL1}
	\boxed{
	\SINR_{k}^{\tt{UL}}
	=
	\frac
	{ p_k\pmb{u}_k^H \pmb{h}_{t,k}\pmb{h}^H_{t,k} \pmb{u}_k }
	{\sum\nolimits_{\substack{k' = 1 \\ k' \neq k }}^K p_{k'} \pmb{u}_k^H \pmb{h}_{t,k'}\pmb{h}^H_{t,k'} \pmb{u}_k  +
	\pmb{u}_k^H \pmb{C}\pmb{Q}\pmb{C}^H \pmb{u}_k + 
	\pmb{u}_k^H \pmb{R}_c \pmb{u}_k +
	\sigma_n^2 \pmb{u}_k^H\pmb{u}_k},
	}
\end{equation}	
 \end{figure*}
In \eqref{eq:SINR_UL1}, $\pmb{Q} = \pmb{V}\pmb{V}^H + \pmb{W}$. Similarly, we can define the achievable rate of transmission of the $k^{th}$ \ac{UL} communication user as
\begin{equation}
\label{eq:rate-equation-2}
	R_{k}^{\tt{UL}} = \log_2( 1 + \SINR_{k}^{\tt{UL}} ), \quad \forall k = 1 \ldots K
\end{equation}
\subsubsection{Secrecy Rates}
Before we clearly express the \textit{secrecy rates}, it is important to define the sum-rates per eavesdropper. First, we define a receive \ac{SINR} between the $p^{th}$ eavesdropper and the $k^{th}$ \ac{UL} communication user as follows

\begin{equation}
	\SINR_{p,k}^{\tt{Eve}}
	=
	\frac
	{\mathbb{E}[\vert g_{k,p} v_k \vert^2]}
	{\sum\limits_{\substack{k' = 1 \\ k' \neq k }}^K \mathbb{E}[\vert g_{k,p} v_k \vert^2] + \mathbb{E}[\vert \alpha_p
	\pmb{a}_{N_T}^H(\theta_p)
	\pmb{x} \vert^2] + \mathbb{E}[\vert z_p \vert^2]},
\end{equation}
which can be re-written as
\begin{equation}
\boxed{
	\SINR_{p,k}^{\tt{Eve}}
	=
	\frac
	{ p_k \vert g_{k,p} \vert^2  }
	{\sum\limits_{\substack{k' = 1 \\ k' \neq k }}^K p_{k'} \vert g_{k',p} \vert^2 + \vert \alpha_p \vert^2 \pmb{a}_{N_T}^H(\theta_p)\pmb{Q}\pmb{a}_{N_T}(\theta_p) + \sigma_p^2}.
	}
\end{equation}

Similarly, we can express the receive \ac{SINR} between the $p^{th}$ eavesdropper and the \ac{DFRC} \ac{BS},
\begin{equation}
	\SINR_{p,\tt{DFRC}}^{\tt{Eve}}
	=
	\frac
	{\mathbb{E}[\vert \alpha_p
	\pmb{a}_{N_T}^H(\theta_p)
	\pmb{Vs} \vert^2]}
	{\sum\limits_{\substack{k = 1  }}^K \mathbb{E}[\vert g_{k,p} v_k \vert^2] + \mathbb{E}[\vert \alpha_p
	\pmb{a}^H_{N_T}(\theta_p)
	\pmb{w} \vert^2] + \mathbb{E}[\vert z_p \vert^2]},
\end{equation}
which gives
\begin{equation}
\boxed{
	\SINR_{p,\tt{DFRC}}^{\tt{Eve}}
	=
	\frac
	{\vert \alpha_p \vert^2 \pmb{a}_{N_T}^H(\theta_p)\pmb{VV}^H\pmb{a}_{N_T}(\theta_p)}
	{\sum\limits_{\substack{k = 1  }}^K p_k \vert g_{k,p} \vert^2 + 
	\vert \alpha_p \vert^2 \pmb{a}_{N_T}^H(\theta_p)\pmb{W}\pmb{a}_{N_T}(\theta_p)
	 + \sigma_p^2 },
	 }
\end{equation}
Now that we have defined the \ac{SINR} expressions per eavesdropper, we take a step further to define the \textit{worst-case secrecy rates} \cite{6119232,6626661,6820768} of the system. Since the eavesdroppers are exposed to \ac{UL} communication signals, as well as \ac{DL} \ac{DFRC} signals, we can define two types of worst-case secrecy rates, i.e.
\begin{equation}
	\label{eq:SR_DL}
	\boxed{
	\SR_{\tt{DL}}
	=
	\min_{\ell,p}
	\Big[
	\log_2(1+\SINR_{\ell}^{\tt{DL}})
	-
	\log_2(1+\SINR_{p,\tt{DFRC}}^{\tt{Eve}})
	\Big]^+,}
\end{equation}
reflecting the worst-case secrecy rate of the system for \ac{DL} communications, and
\useshortskip 
\begin{equation}
	\label{eq:SR_UL}
	\boxed{
	\SR_{\tt{UL}}
	=
	\min_{k,p}
	\Big[
	\log_2(1+\SINR_{k}^{\tt{UL}})
	-
	\log_2(1+\SINR_{p,k}^{\tt{Eve}})
	\Big]^+,}
\end{equation}
reflecting the worst-case secrecy rate of the system for \ac{UL} communications.
\subsubsection{Radar Metrics}
The \ac{DFRC} \ac{BS} aims at performing beam management with each of its transmit vector $\pmb{x}$ in order to maintain highly directional radar transmissions, which are illuminated towards the $P$ eavesdroppers, so as to extract wireless environmental features simultaneously. This allows the \ac{DFRC} \ac{BS} to learn physical attributes related to each one of the eavesdroppers. 

In this article, we adopt the \ac{ISMR} \ac{KPI} \cite{7126203,9287840} reporting radar performance of the transmit beampattern. In fact, this ratio measures sidelobe energy to mainlobe energy as follows
\useshortskip
\begin{equation}
	\ISMR = \frac
	{\int\nolimits_{\pmb{\Theta}_s}\pmb{a}_{N_T}^H(\theta) \pmb{R}_{\pmb{xx}} \pmb{a}_{N_T}(\theta) \ d\theta }
	{\int\nolimits_{\pmb{\Theta}_m}\pmb{a}_{N_T}^H(\theta) \pmb{R}_{\pmb{xx}} \pmb{a}_{N_T}(\theta) \ d\theta }.
\end{equation}
where $\pmb{\Theta}_s$ and $\pmb{\Theta}_m$ are sets containing the directions toward the side-lobe and main-lobe components, respectively. Moreover, $\pmb{R}_{\pmb{xx}}$ represents the transmit covariance matrix. Indeed, using equation \eqref{eq:Tx-signal-BS}, we can write
\begin{equation}
	\pmb{R}_{\pmb{xx}}
	=
	\mathbb{E}(\pmb{x}\pmb{x}^H)
	=
	\pmb{V}\pmb{V}^H
	+
	\pmb{W}.
\end{equation}
The \ac{ISMR} could be reformulated after integrating over directions as
\useshortskip
\begin{equation}
	\ISMR = \frac
	{ \trace\big( \pmb{R}_{\pmb{xx}} \pmb{A}_s \big) }
	{\trace\big( \pmb{R}_{\pmb{xx}} \pmb{A}_m \big)},
\end{equation}
where $\pmb{A}_{s /m} = \int\nolimits_{\pmb{\Theta}_{s/m}} \pmb{a}_{N_T}(\theta)\pmb{a}_{N_T}^H(\theta) \ d\theta  $. In this paper, we choose $P$ mainlobes to be centered around $\theta_1 \ldots \theta_P$, respectively. The lower and upper bounds of the mainlobes are 
\useshortskip
\begin{equation}
	\label{eq:Mainlobe-expression}
	\Theta_m = \bigcup\nolimits_{p=1}^{P} \big\lbrace [\theta_p - \theta_p^{\tt{low}}; \theta_p - \theta_p^{\tt{high}} ] \big\rbrace.
\end{equation}
The \ac{ISMR} metric aims at concentrating power in the mainlobe region(s), while lowering the power in the sidelobe(s) as much as possible to suppress undesired echo bounces, such as clutter and signal-dependent interference.
In optimizing the beams around the target \acp{AoA}, while simultaneously attenuating clutter returns, the signal-clutter-noise ratio is improved. 
As a result, according to \cite{9724174}, the echo return can then be reliably used to estimate additional sensing parameters, such as range based on \ac{ToA} information and speed via doppler information.
In addition to the supplementary sensing information we can estimate, the \ac{ISMR} can then be adapted for tracking the targets by integrating a sophisticated tracking method to carefully steer the mainlobes defined through the \ac{ISMR}.
Therefore, the \ac{ISMR} metric is well-suited for target localization applications and tracking. 
On the other hand, decreasing sidelobe powers in certain directions is, in some applications, desired in order to attenuate unwanted returns, such as interferers or clutter components. A particular and practical instance of \ac{ISMR} appeared in \cite{wasiewicz2015method}, where the concept of \ac{ISMR} is indirectly leveraged. In particular, a radar beamformer is optimized to receive radar signals containing clutter, where the beamformed signal enjoys a high mainlobe and a spatial sidelobe null in the direction of the clutter. 
But even more, the \ac{ISMR} allows flexibility in the choice of how wide the mainlobes can get, which can account for uncertainty on the eavesdropper locations. 
More specifically, the uncertainty on the $p^{th}$ eavesdropper \ac{AoA} can be accomodated by appropriately configuring $\theta_p^{\tt{low}}$ and $\theta_p^{\tt{high}}$.
It is worth noting that the \ac{ISMR} has also been adopted in several works, i.e. \cite{10130609}, \cite{8756034}, and \cite{10018908}.
It is worthwhile pausing here and observing that we are faced with a \textit{"push-pull"} type problem. On one hand, the \ac{DFRC} \ac{BS} aims at probing for energy focusing, in order to extract physical sensing parameters of the eavesdroppers by illuminating simultaneous radar beams centered around locations $\theta_1 \ldots \theta_P$. 
But, on the other hand, the \ac{DFRC} \ac{BS} also aims at revealing minimal communication information towards the malicious eavesdroppers through proper \ac{AN} covariance design. 

\vspace{-0.2cm}
\section{Optimization Framework for Secure \ac{ISAC}}
\label{sec:opt-probs}
In this section, we devise an \ac{ISAC} optimization problem intended for total power minimization that is to be invested on the \ac{UL} power allocation, \ac{DL} beamforming, as well as \ac{AN} generation. We would like to mention that, as highlighted by \cite{5692899}, the ideal case where perfect \ac{CSI} is available at the transmitter, the achievable secrecy rate can be made arbitrarily large by increasing the transmission power. From a power efficiency perspective, though, we intend to find the best power allocation strategy satisfying the \ac{UL} and \ac{DL} achievable secrecy rates. However, to accommodate for \ac{ISAC} and secrecy rate constraints, power minimization takes place under \ac{UL} and \ac{DL} secrecy rates, while guaranteeing a maximum \ac{ISMR}. Based on this, we propose the following optimization problem,
\begin{equation}
 \label{eq:problem1}
\begin{aligned}
(\mathcal{P}_1):
\begin{cases}
\min\limits_{\lbrace \substack{ \pmb{U},
									 \pmb{V},  
									 \pmb{W}, 
									  \pmb{p} }  \rbrace}&   \trace(\pmb{VV}^H) + \trace(\pmb{W}) + \sum\nolimits_{k=1}^K p_k\\
\textrm{s.t.}
 &  \SR_{\tt{DL}} \geq \rho^{\tt{DL}},  \\
&  \SR_{\tt{UL}} \geq  \rho^{\tt{UL}}, \\
& \ISMR \leq \ISMR_{\tt{max}} , \\
 & \pmb{W} \succeq \pmb{0}, \ \pmb{p} \succeq \pmb{0}.   \\
\end{cases}
\end{aligned}
\end{equation}
For convenience, $\pmb{U} \in \mathbb{C}^{N_R \times K}$ contains $\pmb{u}_k$ in its $k^{th}$ column and $\pmb{p}$ is a power vector containing $p_k$ in its $k^{th}$ entry. We first discuss the optimization problem along with its constraints before solving the problem. In particular, we aim at economizing the total transmit power that is to-be invested in \ac{DFRC} operation, \ac{AN}-noise, as well as total power transmitted in the \ac{UL} over all users. Furthermore, we target a minimum acceptable \ac{UL}/\ac{DL} worst-case secrecy rate, captured by the first two constraints, which reflects the amount of information-leakage towards the malicious eavesdroppers. In this context, $\rho^{\tt{DL}}$ and $\rho^{\tt{UL}}$ denote predefined tolerance thresholds for the worst-case secrecy rates in the \ac{DL} and \ac{UL}, respectively. Moreover, the optimization should take place under a maximal \ac{ISMR}, defined hereby through the parameter $\ISMR_{\tt{max}}$, in order to sustain proper radar beams towards the eavesdroppers. This shall guarantee minimum communication information revealed to the eavesdroppers, relative to the \ac{DFRC} \ac{BS} and the \ac{DL} users, while at the same time illuminating beams towards the former. Due to the non-convexity of the secrecy rate constraints, we cast problem \eqref{eq:problem1} as follows
\useshortskip
\begin{equation}
 \label{eq:problem1_1}
\begin{aligned}
(\mathcal{P}_{1.1}):
\begin{cases}
\min\limits_{\lbrace \substack{ \pmb{U},
									 \pmb{V},  
									 \pmb{W}, 
									  \pmb{p} }  \rbrace}&   \trace(\pmb{VV}^H) + \trace(\pmb{W}) + \pmb{1}^T \pmb{p} \\
\textrm{s.t.}
 &  \SINR_{\ell}^{\tt{DL}} \geq \zeta_{\ell}^{\tt{DL}}, \quad \forall \ell, \\
&  \SINR_{k}^{\tt{UL}} \geq  \zeta_{k}^{\tt{UL}}, \quad \forall k, \\
&  \SINR_{p,\tt{DFRC}}^{\tt{Eve}} \leq  \zeta_{p,\tt{DFRC}}^{\tt{Eve}} , \quad \forall p, \\
&  \SINR_{p,k}^{\tt{Eve}} \leq \zeta_{p,k}^{\tt{Eve}} , \quad \forall p,k \\
& \ISMR \leq \ISMR_{\tt{max}} , \\
 & \pmb{W} \succeq \pmb{0}, \  \pmb{p} \succeq \pmb{0},  \\
\end{cases}
\end{aligned}
\end{equation}
where now we define $\zeta_{\ell}^{\tt{DL}}$ and $\zeta_{k}^{\tt{UL}}$ as the minimum accepted \ac{SINR} of the $\ell^{th}$ \ac{DL} and $k^{th}$ \ac{UL} communication users, respectively. Meanwhile, the maximum accepted \ac{SINR} of the $p^{th}$ eavesdropper relative to the \ac{DFRC} is $\zeta_{p,\tt{DFRC}}^{\tt{Eve}}$, and the maximum accepted \ac{SINR} of the $p^{th}$ eavesdropper relative to the $k^{th}$ \ac{UL} communication user is $\zeta_{p,k}^{\tt{Eve}}$. To see the equivalence between the two problems in equations \eqref{eq:problem1} and  \eqref{eq:problem1_1}, we define $LP$ positive real numbers of the form 
$\lbrace \rho^{\tt{DL}}_{\ell,p}  \rbrace_{\substack{\ell = 1 \ldots L\\ p = 1 \ldots P}}$ taking the form of 
\useshortskip
\begin{equation}
\label{eq:SR-decomposition_1}
	\rho^{\tt{DL}}_{\ell,p} = \log_2(1+\zeta_{\ell}^{\tt{DL}}) - \log_2(1+\zeta_{p,\tt{DFRC}}^{\tt{Eve}} ), \quad \forall  \ell , p
\end{equation}
where $\min\nolimits_{\ell,p} \rho^{\tt{DL}}_{\ell,p} = \rho^{\tt{DL}}$ and $\zeta_{\ell}^{\tt{DL}} > \zeta_{p,\tt{DFRC}}^{\tt{Eve}}$. 
Next, using the definition of \ac{DL} worst-case secrecy rate in equation \eqref{eq:SR_DL} on the first constraint of problem \eqref{eq:problem1}, we get
\useshortskip
\begin{equation}
\label{eq:SR-bound_1}
\log_2(1+\SINR_{\ell}^{\tt{DL}})
-
\log_2(1+\SINR_{p,\tt{DFRC}}^{\tt{Eve}})
\geq
 \rho^{\tt{DL}}_{\ell,p}, \quad \forall  \ell , p
\end{equation}
Finally, using \eqref{eq:SR-decomposition_1} in \eqref{eq:SR-bound_1}, we can satisfy the first constraint in \eqref{eq:problem1} 
by upper bounding $\SINR_{\ell}^{\tt{DL}}$ through the first constraint of \eqref{eq:problem1_1},
and lower bounding $\SINR_{p,\tt{DFRC}}^{\tt{Eve}}$ through the third constraint of \eqref{eq:problem1_1}.
A similar argument is used by utilizing the definition of the \ac{UL} worst-case secrecy rate in equation \eqref{eq:SR_UL} to arrive at the second and fourth constraints appearing in problem \eqref{eq:problem1_1}. In the following section, we propose an iterative method to solve \eqref{eq:problem1_1}.


\vspace{-0.2cm}
\section{Secure \ac{ISAC} via Successive Convex Approximations}
\label{sec:opt-01}
\subsection{Algorithmic Derivation}
\label{subsec:algo-derivation}
We start by optimizing with respect to $\pmb{U}$. Note that the only quantities depending on $\pmb{U}$ are $\lbrace \SINR_{k}^{\tt{UL}}  \rbrace_{k=1}^K$. Therefore, we first aim to maximize all the \ac{UL} \acp{SINR} with respect to $\pmb{U}$. Based on this, we consider
\useshortskip
\begin{equation}
	\label{eq:Opt-wrt-U}
	\Big\lbrace
	\widehat{\pmb{U}}
	=
	\argmax_{\pmb{U}} \sum\limits_{k=1}^K \SINR_{k}^{\tt{UL}} 
	\Big\rbrace
	\propto
	\Big\lbrace \widehat{\pmb{u}}_k = \argmax_{\pmb{u}_k}  \SINR_{k}^{\tt{UL}} \Big\rbrace.
\end{equation}
Note that $\SINR_{k}^{\tt{UL}}$ forms a generalized Rayleigh quotient in $\pmb{u}_k$ taking the form of
\useshortskip
\begin{equation}
	\label{eq:SINR_UL1_rayliegh}
\begin{split}
	&\SINR_{k}^{\tt{UL}}
	\\ & =
	\frac
	{ p_k \pmb{u}_k^H \pmb{h}_{t,k}\pmb{h}^H_{t,k} \pmb{u}_k }
	{\pmb{u}_k^H \Bigg(\sum\nolimits_{\substack{k' = 1 \\ k' \neq k }}^K  p_{k'} \pmb{h}_{t,k'}\pmb{h}^H_{t,k'}   +
	 \pmb{C}\pmb{Q}\pmb{C}^H + 
	 \pmb{R}_c +
	\sigma_n^2 \pmb{I} \Bigg) \pmb{u}_k}. \quad \forall k 
\end{split}
\end{equation}
Following \cite{7872443}, the solution of the optimization problem in \eqref{eq:Opt-wrt-U} is readily obtained via
\useshortskip
\begin{equation}
\small
\begin{split}
	\widehat{\pmb{u}}_k 
	\propto
	\max \eigvec 
	\Bigg[
	 \Bigg(\sum\limits_{\substack{k' = 1 \\ k' \neq k }}^K & p_{k'}\pmb{h}_{t,k'}\pmb{h}^H_{t,k'}   +
	 \pmb{C}\pmb{Q}\pmb{C}^H + 
	 \pmb{R}_c +
	\sigma_n^2 \pmb{I} \Bigg)^{-1} \\ &
	 \pmb{h}_{t,k}\pmb{h}^H_{t,k}
	 \Bigg], \forall k = 1 \ldots K.
\end{split}
\end{equation}
But since $\pmb{h}_{t,k}\pmb{h}^H_{t,k}$ is $\rank$-one, then the optimal beamforming vector maximizing the \ac{SINR} at the $k^{th}$ \ac{UL} communication user can be expressed as
\useshortskip
\begin{equation}
	\label{eq:uk-in-closed-form}
	\widehat{\pmb{u}}_k 
	=
	\Bigg(\sum\nolimits_{\substack{k' = 1 \\ k' \neq k }}^K  p_{k'}\pmb{h}_{t,k'}\pmb{h}^H_{t,k'}   +
	 \pmb{C}\pmb{Q}\pmb{C}^H + 
	 \pmb{R}_c +
	\sigma_n^2 \pmb{I} \Bigg)^{-1} 
	 \pmb{h}_{t,k},
\end{equation}
Treating $\widehat{\pmb{u}}_k$ as nuisance parameters, we plug their expressions back in \eqref{eq:problem1_1} to get a problem independent of $\pmb{U}$ as
\useshortskip
\begin{equation}
 \label{eq:problem1_2}
\begin{aligned}
(\mathcal{P}_{1.2}):
\begin{cases}
\min\limits_{\lbrace \substack{     \pmb{V},  
									 \pmb{W}, 
									  \pmb{p} }  \rbrace}&   \sum\nolimits_{\ell = 1}^L\trace(\pmb{\pmb{V}}_\ell) + \trace(\pmb{W}) + \pmb{1}^T \pmb{p} \\
\textrm{s.t.}
 &  \SINR_{\ell}^{\tt{DL}} \geq \zeta_{\ell}^{\tt{DL}}, \quad \forall \ell, \\
&  \widehat{\SINR}_{k}^{\tt{UL}} \geq  \zeta_{k}^{\tt{UL}}, \quad \forall k, \\
&  \SINR_{p,\tt{DFRC}}^{\tt{Eve}} \leq  \zeta_{p,\tt{DFRC}}^{\tt{Eve}} , \quad \forall p, \\
&  \SINR_{p,k}^{\tt{Eve}} \leq \zeta_{p,k}^{\tt{Eve}} , \quad \forall p,k \\
& \ISMR \leq \ISMR_{\tt{max}} , \\
 & \pmb{W} \succeq \pmb{0},  \ \pmb{p} \succeq \pmb{0},\ \pmb{V}_\ell \succeq \pmb{0}, \quad \forall \ell    \\
 & \rank(\pmb{V}_\ell) = 1, \quad \forall \ell
\end{cases}
\end{aligned}
\end{equation}
where the \ac{UL} \ac{SINR} of the $k^{th}$ user after beamforming as seen by the \ac{DFRC} \ac{BS} is given as
\useshortskip 
\begin{equation}
\begin{split}
	&\widehat{\SINR}_{k}^{\tt{UL}}
	\\&= 
	 p_k
	 \pmb{h}^H_{t,k}
	\Bigg(\sum\limits_{\substack{k' = 1 \\ k' \neq k }}^K  p_{k'}\pmb{h}_{t,k'}\pmb{h}^H_{t,k'}   +
	 \pmb{C}\pmb{Q}\pmb{C}^H + 
	 \pmb{R}_c +
	\sigma_n^2 \pmb{I} \Bigg)^{-1} 
	 \pmb{h}_{t,k}.
\end{split}
\end{equation}
Note that we have replaced the term $\trace(\pmb{VV}^H)$ by $\sum\nolimits_{\ell = 1}^L\trace(\pmb{\pmb{V}}_\ell)$, where each $\pmb{\pmb{V}}_\ell$ is constrained to be rank-one.
The first constraint in \eqref{eq:problem1_2} is written in terms of traces as,
\useshortskip
\begin{equation}
	\label{eq:constraint-one-with-UL-interference}
\begin{split}
	\trace \lbrace \big(\frac{1}{\zeta_{\ell}^{\tt{DL}}}\pmb{V}_\ell -  
	(\sum\limits_{\substack{\ell^{'} \neq \ell}}^L\pmb{V}_{\ell'} + \pmb{W} )
	\big)\pmb{H}_{r,\ell} \rbrace
	\geq \Big( \sum\limits_{\substack{k = 1  }}^K p_k \vert q_{k,\ell} \vert^2 
	+ 
	\sigma_{\ell}^2
	\Big),
\end{split}
\end{equation}
where $\pmb{H}_{r,\ell} = \pmb{h}_{r,\ell} \pmb{h}_{r,\ell}^H$, which describes a convex set. 
The second constraint in \eqref{eq:problem1_2} is non-convex. To remedy this non-convexity, we resort to the \ac{SCA} technique to relax the non-convex constraint. Thus, via sequential convex programming in an iterative fashion, we can provide a locally optimal solution. Before we proceed, we write the second constraint in \eqref{eq:problem1_2} as 
\begin{equation}
	 p_k\pmb{h}^H_{t,k}
	\pmb{\Phi}_k^{-1}(\pmb{Q},\pmb{p})
	 \pmb{h}_{t,k} \geq \zeta_{k}^{\tt{UL}} , \quad \forall k = 1 \ldots K,
\end{equation}
where $\pmb{\Phi}_k(\pmb{Q},\pmb{p}) = \sum\nolimits_{\substack{k' = 1 \\ k' \neq k }}^K  p_{k'}\pmb{h}_{t,k'}\pmb{h}^H_{t,k'}   +
	 \pmb{C}\pmb{Q}\pmb{C}^H + 
	 \pmb{R}_c +
	\sigma_n^2 \pmb{I}.$
Now, exploring the following first-order Taylor series approximation about a given estimate of $\pmb{\Phi}_k^{-1}(\pmb{Q},\pmb{p})$, say $ [\pmb{\Phi}_k^{(m)}]^{-1} \triangleq \pmb{\Phi}_k^{-1}(\pmb{Q}^{(m)},\pmb{p}^{(m)})$, which is the obtained estimate of $\pmb{\Phi}_k^{-1}(\pmb{Q},\pmb{p})$ at the $m^{th}$ iteration of the \ac{SCA} method,
\useshortskip
\begin{equation}
\begin{split}
	& \pmb{h}^H_{t,k}
	\pmb{\Phi}_k^{-1}(\pmb{Q},\pmb{p})
	 \pmb{h}_{t,k} 
	 \geq
	 \pmb{h}^H_{t,k}
	[\pmb{\Phi}_k^{(m)}]^{-1}
	 \pmb{h}_{t,k}
	 \\ & - \pmb{h}^H_{t,k}
	[\pmb{\Phi}_k^{(m)}]^{-1}
	\big( \pmb{\Phi}_k -  \pmb{\Phi}_k^{(m)}\big)
	[\pmb{\Phi}_k^{(m)}]^{-1}
	\pmb{h}_{t,k}.
\end{split} 
\end{equation}
Using the above lower-bound to upper-bound $\zeta_{k}^{\tt{UL}}$ for all $k = 1 \ldots K$, we can write at iteration $m$ the following
\useshortskip
\begin{equation}
\label{eq:constraint38}
\begin{split} 
	  \pmb{\Lambda}_k \leq 0
	  ,
	  \
	  \forall k = 1 \ldots k.
\end{split}
\end{equation}
where $ \pmb{\Lambda}_k = 
	\frac{\zeta_{k}^{\tt{UL}}}{p_k}
	  +
	  \pmb{h}^H_{t,k}
	[\pmb{\Phi}_k^{(m)}]^{-1}
	\pmb{\Phi}_k
	[\pmb{\Phi}_k^{(m)}]^{-1}
	\pmb{h}_{t,k}
	-
	2  \pmb{h}^H_{t,k}
	[\pmb{\Phi}_k^{(m)}]^{-1}
	 \pmb{h}_{t,k}$.
The first order derivatives of $\pmb{\Lambda}_k$ are
\begin{align*}
	\frac{\partial \pmb{\Lambda}_k}{\partial p_k}
	&=
	-
	\frac{\zeta_{k}^{\tt{UL}}}{p_k^2},\\
	\frac{\partial \pmb{\Lambda}_k}{\partial p_{k'}}
	&=
	\pmb{h}^H_{t,k}
	[\pmb{\Phi}_k^{(m)}]^{-1}
	\pmb{h}_{t,k'}\pmb{h}_{t,k'}^H
	[\pmb{\Phi}_k^{(m)}]^{-1}
	\pmb{h}_{t,k}, \quad \forall k' \neq k, \\
	\frac{\partial \pmb{\Lambda}_k}{\partial \pmb{V}_\ell}	
	&=
	\frac{\partial \pmb{\Lambda}_k}{\partial \pmb{W}}
	=
	\pmb{C}^H
	[\pmb{\Phi}_k^{(m)}]^{-1}
	\pmb{h}_{t,k}
	\pmb{h}_{t,k}^H
	[\pmb{\Phi}_k^{(m)}]^{-1}
	\pmb{C}, \quad \forall \ell.
\end{align*}
As the expression $\pmb{\Lambda}_k$ is twice differentiable, we can see that from the above expressions, all second-order derivatives are zero, except for
\useshortskip
\begin{equation}
	\frac{\partial^2 \pmb{\Lambda}_k}{\partial p_k^2}
	=
	\frac{2\zeta_{k}^{\tt{UL}}}{p_k^3}.
\end{equation}
This means that the Hessian $\nabla^2 \pmb{\Lambda}_k$ is an all-zero matrix except for $\frac{2\zeta_{k}^{\tt{UL}}}{p_k^3}$ at its $k^{th}$ diagonal entry. 
Therefore, $\nabla^2 \pmb{\Lambda}_k \succeq \pmb{0}$ if and only if $\frac{2\zeta_{k}^{\tt{UL}}}{p_k^3} \geq 0 $, which is valid since the powers $p_k$ are constrained to be positive and $\zeta_{k}^{\tt{UL}}$ is a given positive quantity.
Finally, we conclude that $\pmb{\Lambda}_k$ is convex in $\pmb{p}$, $\pmb{V}_1 \ldots \pmb{V}_L$ and $\pmb{W}$.
Moreover, the third constraint in \eqref{eq:problem1_2} is written as
\useshortskip
\begin{equation}
\begin{split}
	&\frac{1}{\zeta_{p,\tt{DFRC}}^{\tt{Eve}} }
	\vert \alpha_p \vert^2 \pmb{a}_{N_T}^H(\theta_p)\Big( \sum\nolimits_{\ell = 1}^L\trace(\pmb{\pmb{V}}_\ell) \Big)\pmb{a}_{N_T}(\theta_p)
	 \\ & \leq 
	\sum\nolimits_{\substack{k = 1  }}^K p_k \vert g_{k,p} \vert^2 + 
	\vert \alpha_p \vert^2 \pmb{a}_{N_T}^H(\theta_p)\pmb{W}\pmb{a}_{N_T}(\theta_p)
	 + \sigma_p^2,
\end{split}
\end{equation}
where, as done on the objective function, we have replaced $\pmb{VV}^H$ with $\sum\nolimits_{\ell = 1}^L\trace(\pmb{\pmb{V}}_\ell)$.  
Furthermore, the fourth constraint in \eqref{eq:problem1_2} can be expressed as follows
\useshortskip
\begin{equation}
\begin{split}
	& \frac{1}{\zeta_{p,k}^{\tt{Eve}} }
	p_k \vert g_{k,p} \vert^2 
	\\ &  -
	\Big(
	\sum\limits_{\substack{ k' \neq k }}^K p_{k'} \vert g_{k',p} \vert^2 + \vert \alpha_p \vert^2 \pmb{a}_{N_T}^H(\theta_p)\pmb{Q}\pmb{a}_{N_T}(\theta_p) + \sigma_p^2
	\Big) \leq 0.
\end{split}
\end{equation}
Finally, dropping the $\rank-$one constraint in the $m^{th}$ iteration of the \ac{SCA} method, the optimization problem in \eqref{eq:problem1_2} can be translated to \eqref{eq:problem1_3}. As shall be seen in the next section, the $\rank-$one relaxation does not introduce a sub-optimality due to the special structure of the optimization problem in hand.

Note that problem $(\mathcal{P}_{1.3}^{(m)})$ is a convex optimization problem and can be readily solved via classical solvers, e.g., the {\tt{CVX}} toolbox in {\tt{MATLAB}}. Before we summarize our \ac{SCA}-based algorithm, we introduce an optimality theorem related to the optimization problem at hand, which will further aid us at generating the beamforming vectors.
The implicit presence of perfect \ac{CSI} knowledge at all given nodes in an \ac{ISAC} system is a simplifying assumption.
In reality, this knowledge cannot be perfect as inaccurate estimation and quantization errors, as well as outdated channel effects, are part of transmit and receive paths.
The imperfect \ac{CSI} case can be accommodated as part of this design by leveraging the ellipsoidal error model on the available \ac{CSI} vectors.
For example, we can model $\pmb{h}_{r,\ell}
	=
	\widehat{\pmb{h}}_{r,\ell}
	+ 
	\Delta \pmb{h}_{r,\ell}$, where $\Vert 
\pmb{E}_{r,\ell}^{\frac{1}{2}} 
\Delta \pmb{h}_{r,\ell} 
\Vert 
\leq
1$ for all $\ell$. Here, $\pmb{E}_{r,\ell}$ controls the ellipsoidal shape of the \ac{CSI} perturbation set on the $\ell^{th}$ \ac{DL} channel.
A similar model can be adopted for all other links.
Furthermore, a \ac{BCCD} type algorithm can be further devised due to the decoupled nature of some constraints that follow due to the imperfect \ac{CSI}  formulation.
However, due to lack of space, we defer the imperfect \ac{CSI} extension of the proposed secure \ac{FD} \ac{ISAC} problem as part of our future work.

\begin{algorithm}[H]
\caption{\ac{SCA}-based method to optimize \eqref{eq:problem1}}\label{alg:alg1}
\begin{algorithmic}
\STATE 
\STATE {\textbf{Input}: $\lbrace \zeta_{\ell}^{\tt{DL}},\pmb{h}_{r,\ell}, \sigma_\ell \rbrace_{\ell = 1}^L$, $\lbrace\zeta_{k}^{\tt{UL}}, \pmb{h}_{t,k}  \rbrace_{k=1}^K$, $ \lbrace \zeta_{p,\tt{DFRC}}^{\tt{Eve}}, \gamma_p,\theta_p , \sigma_p, \alpha_p \rbrace_{p=1}^P$, $\lbrace \zeta_{p,k}^{\tt{Eve}}, g_{k,p} \rbrace_{p,k=1}^{P,K}$, $\ISMR_{\tt{max}}$, $\beta$, $\pmb{H}_{\tt{SI}}$, $\pmb{A}_s$, $\pmb{A}_m$. }
\STATE \textbf{Initialize}:
\STATE \hspace{0.5cm}  Set $m=0$,  $\pmb{V}_\ell^{(0)}= \pmb{I}$, $\pmb{W}^{(0)} = \pmb{0}$, $\pmb{p}^{(0)} = \frac{1}{K} \pmb{1}$.
\STATE \hspace{0.5cm} Compute $\pmb{C} = \sum\nolimits_{p=1}^P \gamma_p \pmb{a}_{N_R}(\theta_p)\pmb{a}_{N_T}^H(\theta_p) + \sqrt{\beta}\HSI $.
\STATE \textbf{while} $ m < M_{iter}$
\STATE \hspace{0.5cm} Solve $(\mathcal{P}_{1.3}^{(m)})$ in \eqref{eq:problem1_3} to get $\pmb{V}_\ell^{(m+1)},\pmb{W}^{(m+1)}, \pmb{p}^{(m+1)}$.
\STATE \hspace{0.5cm} Update $\pmb{Q}^{(m+1)}$ as $\pmb{Q}^{(m+1)} =\sum\nolimits_{\ell=1}^L \pmb{V}_\ell^{(m+1)} + \pmb{W}^{(m+1)}$.
\STATE \hspace{0.5cm} Update $\pmb{\Phi}_k^{(m+1)}$ for all $k$ as 
\STATE \hspace{0.5cm} $m \gets m + 1$
\STATE \hspace{0.5cm} $
	\pmb{\Phi}_k^{(m)} = \sum\limits_{\substack{k' = 1 \\ k' \neq k }}^K  p_{k'}^{(m)}\pmb{h}_{t,k'}\pmb{h}^H_{t,k'}   +
	 \pmb{C}\pmb{Q}^{(m)}\pmb{C}^H + 
	 \pmb{R}_c +
	\sigma_n^2 \pmb{I}
$.
\STATE \textbf{set} $\widehat{\pmb{Q}} = \pmb{Q}^{(m)}$, $\widehat{\pmb{W}} = \pmb{W}^{(m)}$, $\widehat{\pmb{p}} = \pmb{p}^{(m)}$, $\lbrace \widehat{\pmb{V}}_\ell = {\pmb{V}}_\ell^{(m)} \rbrace_{\ell = 1}^L$.
\STATE \textbf{for} $k = 1 \ldots K$
\STATE \hspace{0.5cm} $\widehat{\pmb{u}}_k 
	=
	\Bigg(\sum\limits_{\substack{k' = 1 \\ k' \neq k }}^K  \widehat{p}_{k'}\pmb{h}_{t,k'}\pmb{h}^H_{t,k'}   +
	 \pmb{C}\widehat{\pmb{Q}}\pmb{C}^H + 
	 \pmb{R}_c +
	\sigma_n^2 \pmb{I} \Bigg)^{-1} 
	 \pmb{h}_{t,k}$.
\STATE  \textbf{for} $ \ell = 1 \ldots L$
\STATE \hspace{0.5cm} $\lambda_\ell \pmb{v}_\ell \pmb{v}_\ell^H \gets\EVD \big( {\pmb{V}}_\ell^{(M_{iter})} \big)$.
\STATE \hspace{0.5cm} $\widehat{\pmb{V}}_{[:,\ell]} \gets \sqrt{\lambda_\ell} \pmb{v}_\ell$.
\STATE \textbf{return}  $\lbrace \widehat{\pmb{u}}_k  \rbrace_{k=1}^K$, $\widehat{\pmb{W}}$, $\widehat{\pmb{p}}$, $ \widehat{\pmb{V}}$.
\end{algorithmic}
\end{algorithm}

\subsection{Optimality}
Ideally, we would aim to produce \ac{DL} $\rank-$one beamforming solutions of the form $\pmb{v}_\ell\pmb{v}_\ell^H$. Otherwise, post-processing procedures are required to produce $\rank$-one approximations of $\pmb{V}_\ell$, $\forall \ell$. Accordingly, Gaussian randomization techniques or principle component analysis methods are commonly employed to obtain a suboptimal solution. These approximations include Gaussian randomization \cite{1634819,10014666}. However, in our case, due to the special structure of the problem, we have the following result.

\textbf{Theorem 1 ($\rank$-one Optimality): } \textit{At the $m^{th}$ iteration of \textbf{Algorithm \ref{alg:alg1}}, consider the optimization problem 
$(\mathcal{P}_{1.3}^{(m)})$ given in equation \eqref{eq:problem1_3}. Then, for all $\ell = 1 \ldots L$, the solution of $(\mathcal{P}_{1.3}^{(m)})$ satisfies }
\begin{equation}
	\rank(\pmb{V}_\ell^{(m)}) = 1, \quad \forall \ell = 1 \ldots L.
\end{equation}
The proof of this theorem is detailed in Appendix \ref{appendix:proof-rank-1}. It turns out that our relaxation is not a "relaxation", after all. Given the $\rank-$one optimality of $\rank(\pmb{V}_1^{(m)}) \ldots \rank(\pmb{V}_L^{(m)})$, an \ac{EVD} suffices to recover $\pmb{V}$. This is achieved as follows
\vspace{-0.2cm}
\begin{equation}
	\pmb{V}_{[:,\ell]}
	=
	\sqrt{\lambda_\ell^{(\infty)}}
	\pmb{v}_\ell^{(\infty)},
\end{equation}
where $\lambda_\ell^{(\infty)}$ and $\pmb{v}_\ell^{(\infty)}$ represent the non-zero eigenvalue and its corresponding eigenvector of the final converged quantity given by \textbf{Algorithm \ref{alg:alg1}}, i.e. $\pmb{V}_\ell^{(\infty)}$, respectively. The above theorem tells us that the $\rank$-one relaxation does not introduce any sort of sub-optimality onto the relaxed problem. An instance of such an optimality appears in \cite{10018908}, where \ac{ISAC} beamforming was used to study outage performances in the case of imperfect channel state information. Furthermore, thanks to this theorem, we can now generate the \ac{DL} beamforming vectors through \ac{EVD}. To this end, the proposed algorithm is summarized in \textbf{Algorithm \ref{alg:alg1}}.


 \begin{figure*}[b]
\input{Actions/Problem1o3.tex}
\end{figure*}


\vspace{-0.2cm}
\section{Simulation Results}
\label{sec:simulations}
In this section, we present our simulation findings. Before we analyze and discuss our results, we mention the simulation parameters and benchmarks used in simulations.
\begin{figure*}[t]
\centering
\subfloat[$\sigma_n^2 = -80\dBm$]{\includegraphics[width=2.5in]{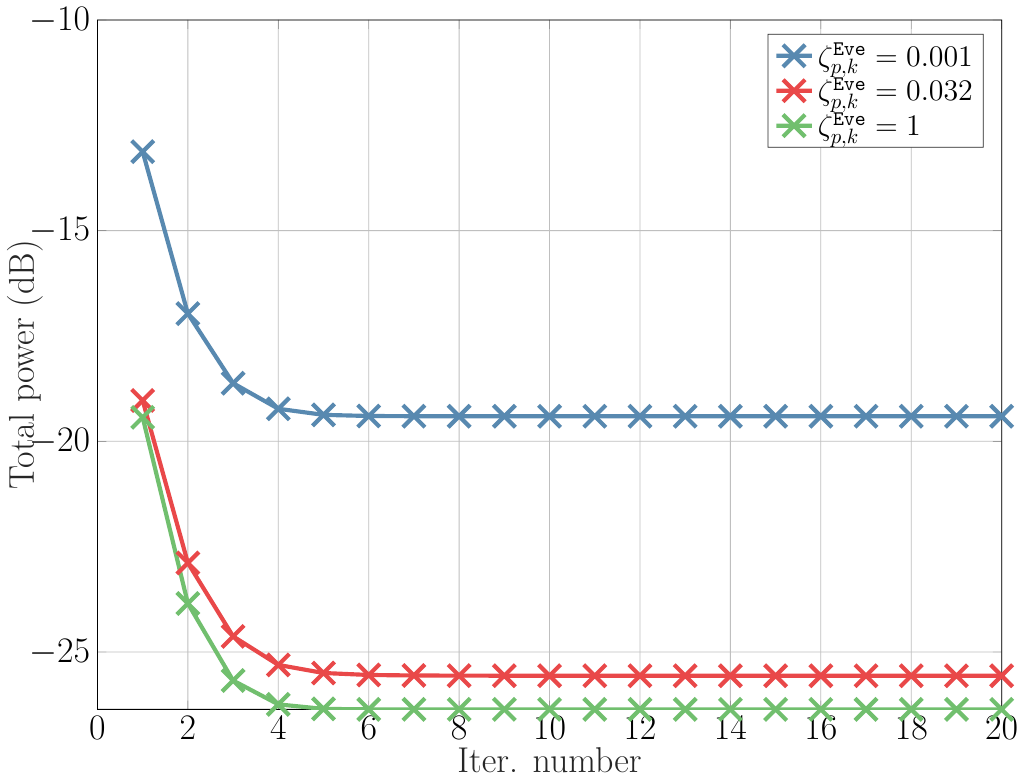}%
\label{fig:pow_consumption_1}}
\hfil
\subfloat[$\sigma_n^2 = -60\dBm$]{\includegraphics[width=2.5in]{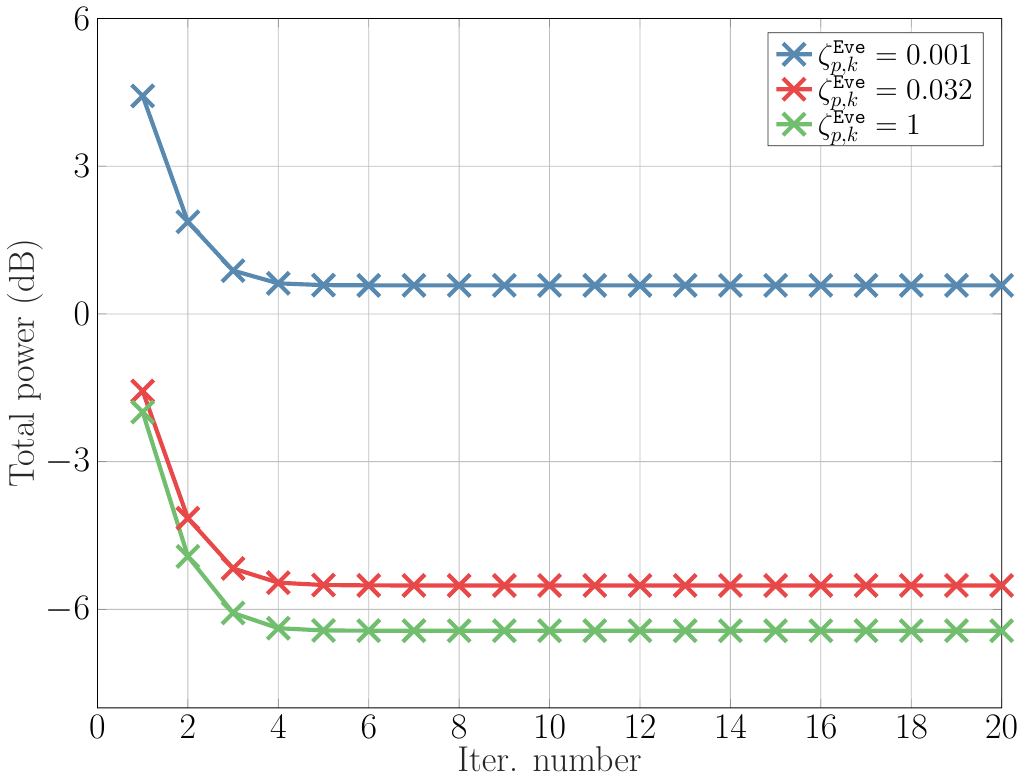}%
\label{fig:pow_consumption_2}}
\caption{\protect\input{Actions/Captions/Fig2a_captions.tex}}
\label{fig:pow_consumption}
\end{figure*}

\begin{figure}[t]
	\centering
	\includegraphics[width=0.75\linewidth]{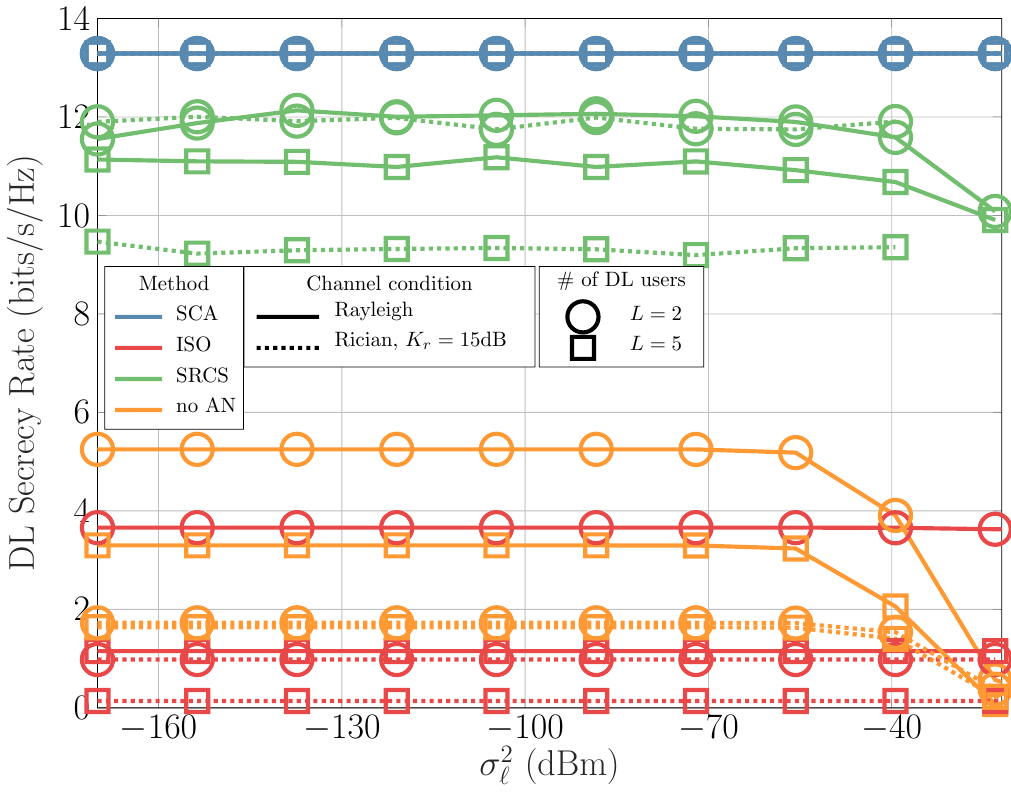}
\caption{\protect\input{Actions/Captions/Fig3_captions.tex}}
	\label{fig:DL_secrecy_1}
\end{figure}

\begin{figure}[t]
	\centering
	\includegraphics[width=0.8\linewidth]{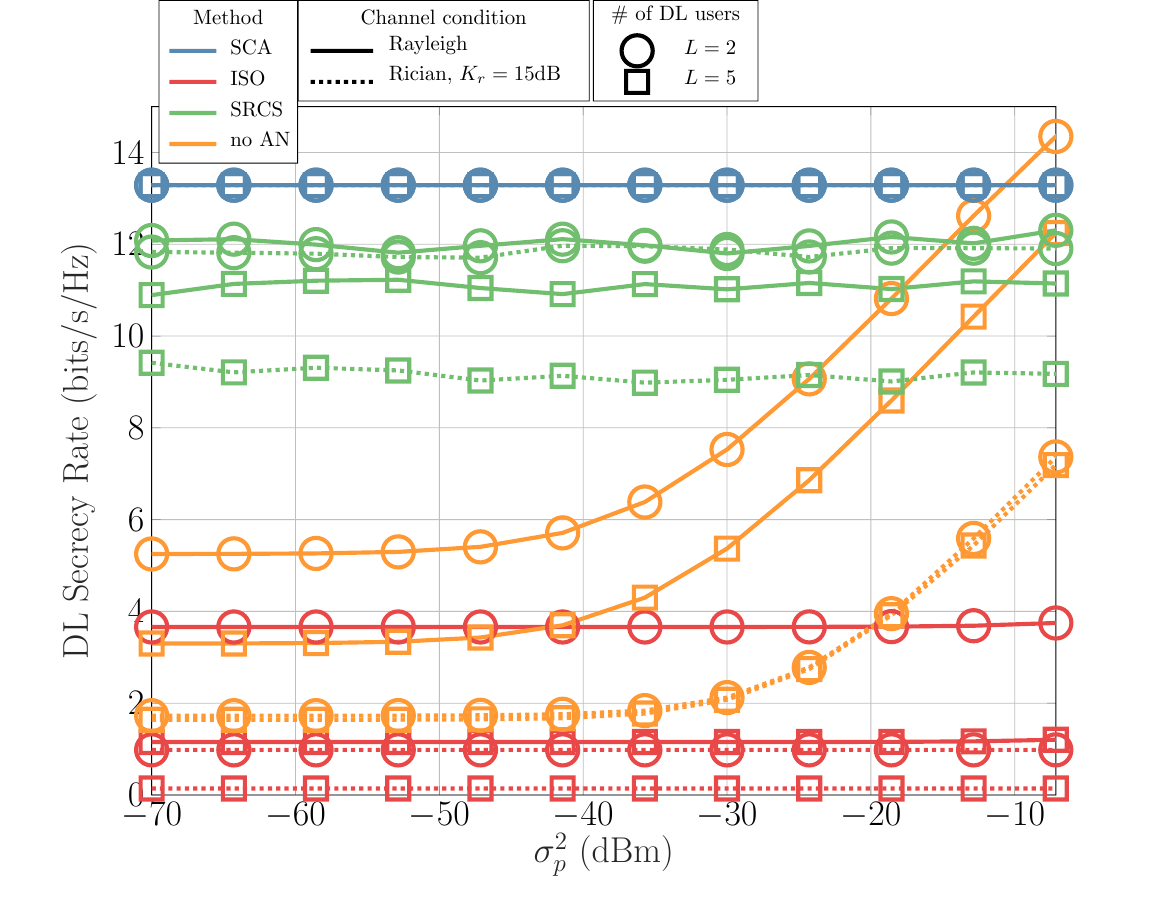}
\caption{\protect\input{Actions/Captions/Fig4_captions.tex}}
	\label{fig:DL_secrecy_2}
\end{figure}

\begin{figure}[t]
	\centering
	\includegraphics[width=0.85\linewidth]{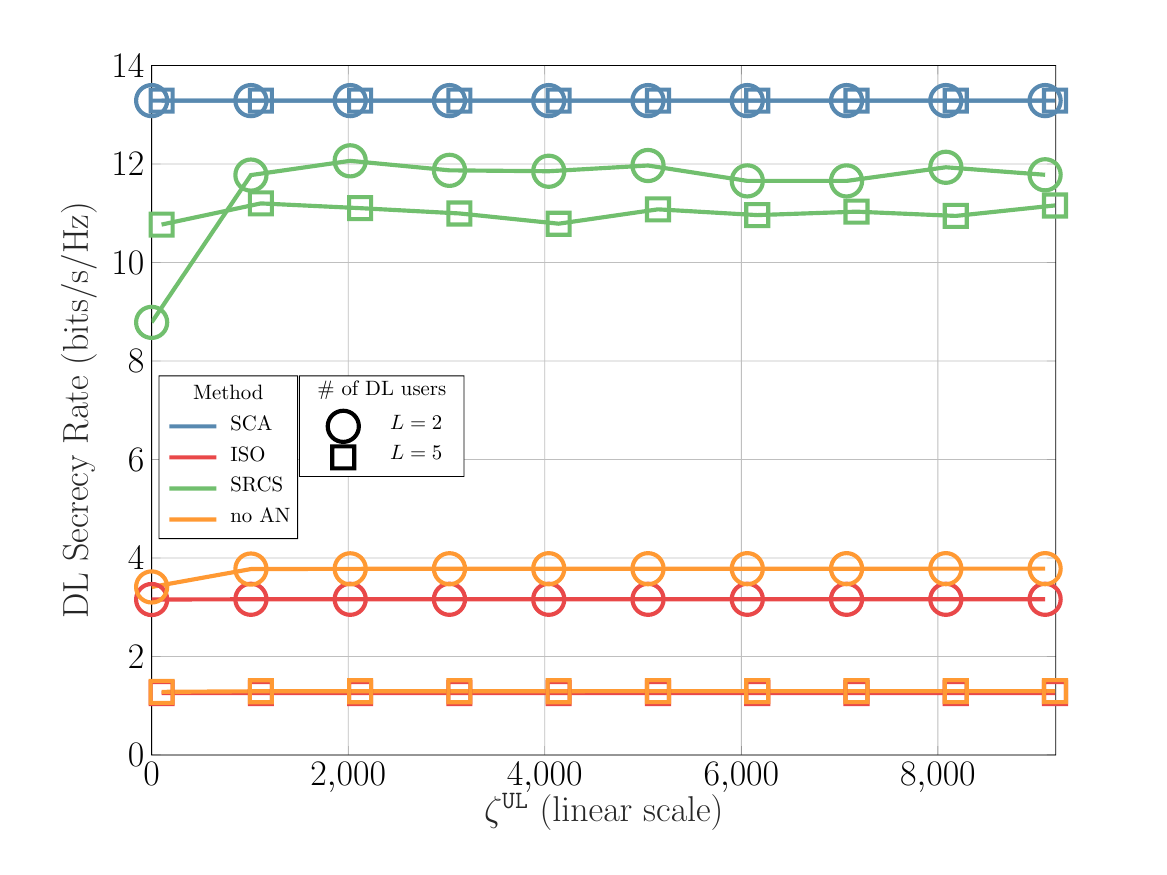}
	\caption{\protect\input{Actions/Captions/Fig5_captions.tex}}
	\label{fig:DL_secrecy_3}
\end{figure}

\begin{figure}[t]
	\centering
	\includegraphics[width=1\linewidth]{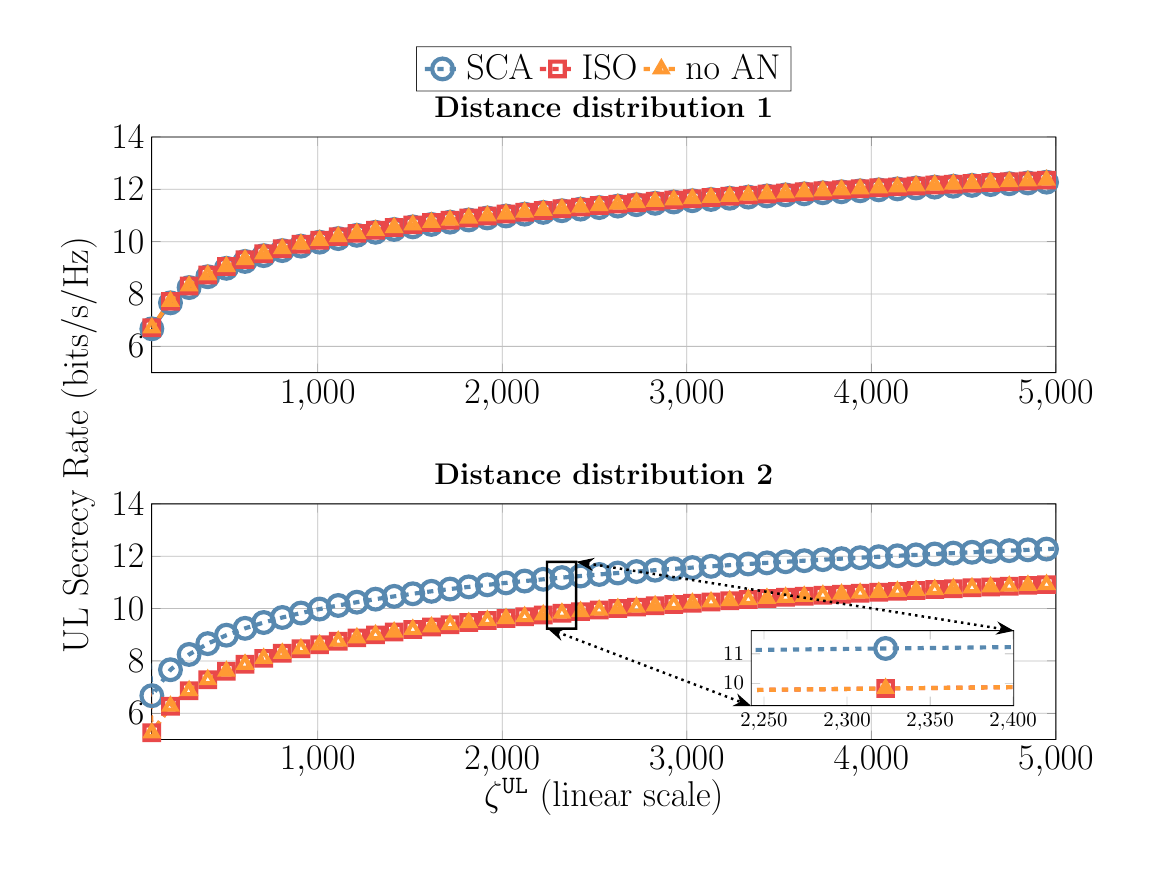}
	\caption{\protect\input{Actions/Captions/Fig6_caption.tex}}
		\label{fig:UL_secrecy_1}
\end{figure}

\begin{figure}[t]
	\centering
	\includegraphics[width=1\linewidth]{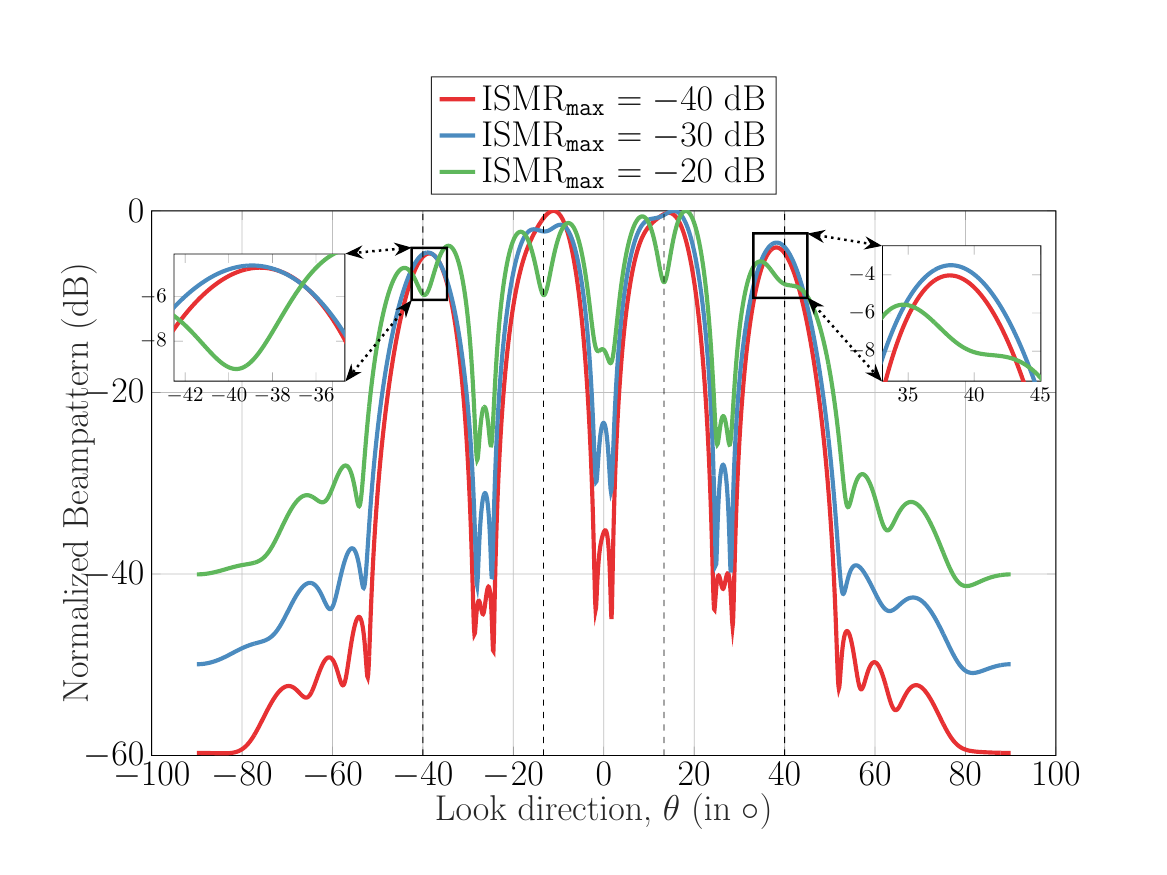}
	\caption{\protect\input{Actions/Captions/Fig7_caption.tex}}
	\label{fig:bp_1}
\end{figure}

\begin{figure}[t]
	\centering
	\includegraphics[width=0.8\linewidth]{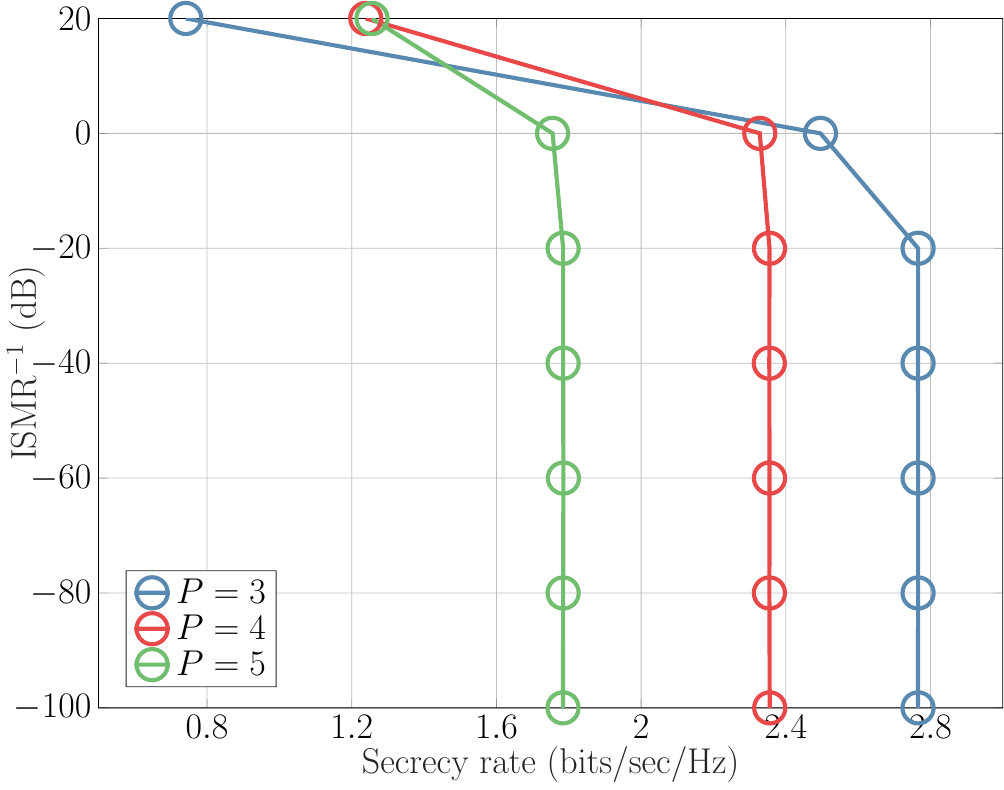}
	\caption{\protect\input{Actions/Captions/Fig8_captions.tex}}
	\label{fig:tradeoff}
\end{figure}

\begin{figure}[t]
	\centering
	\includegraphics[width=1\linewidth]{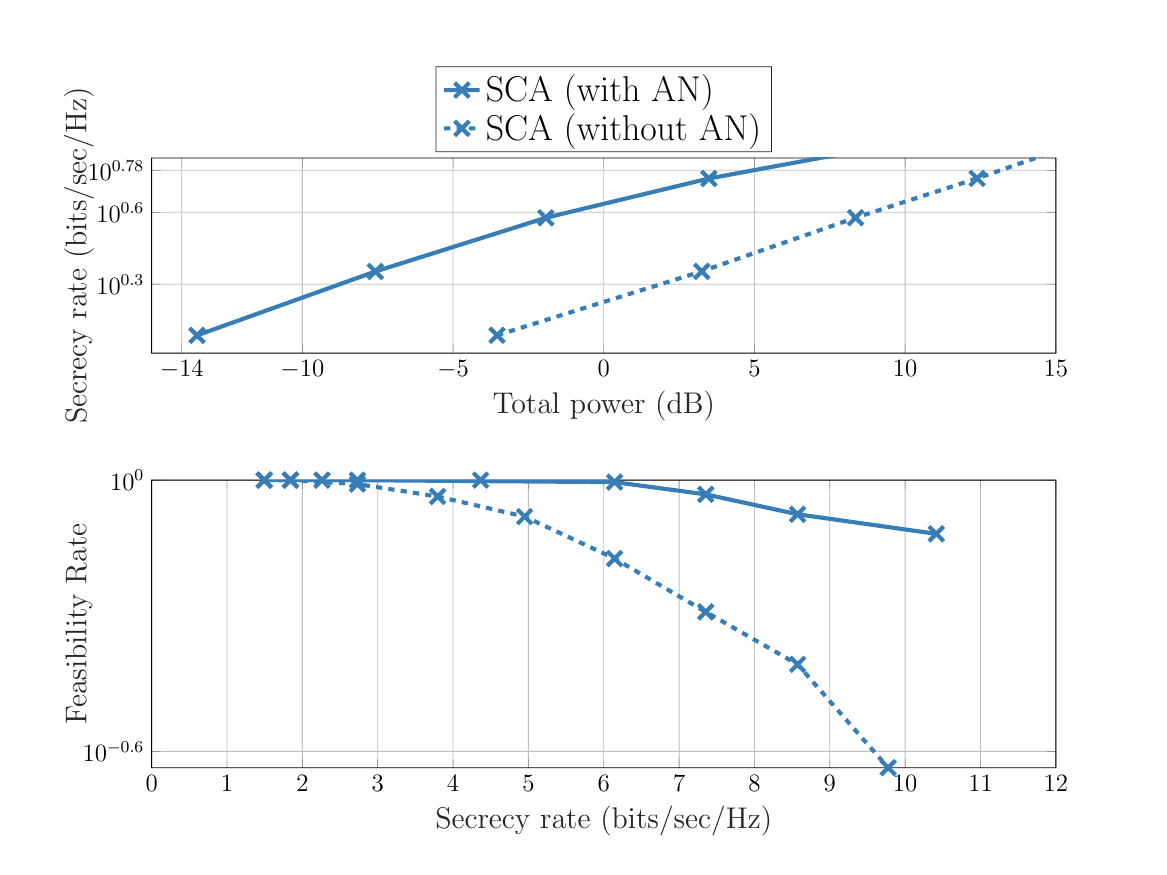}
	\caption{\protect\input{Actions/Captions/Fig9_captions.tex}}
	\label{fig:tradeoff2}
\end{figure}
\vspace{-0.4cm}
\subsection{Parameter Setup}
Unless otherwise stated, we use the simulation parameters depicted in Table \ref{tab:table1}.
Note that we have chosen the users and eavesdroppers to fall within the typical cell radius for \ac{mmWave} communications, which is $500\meters$ \cite{8340227}. 
Monte Carlo type simulations have been performed in order to analyze the efficacy of the proposed secure \ac{ISAC} beamforming and power allocation method.
We consider perfect \ac{CSI} of all nodes in the system, as the case of imperfect \ac{CSI} is left out for future work.
Moreover, the symbols are assumed to be \ac{i.i.d} and independent from noise realizations. 
Furthermore, we use Rayleigh \cite{9217488} and Rician distributions \cite{8207426} to simulate different channel conditions, with log-normal shadowing of $20\dB$ and free space pathloss.
The carrier frequency is centered around $28\GHz$.
The transmit and receive array at the \ac{DFRC} \ac{BS} follow a \ac{ULA} fashion, with $N_T = N_R = 12$ antennas and inter-element spacing of $\frac{\lambda}{2}$. In addition, the transmit and receive antenna gains are $25\dBi$.
The \ac{UL} and \ac{DL} users are equipped with single antennas with antenna gains of $17\dBi$ and $12\dBi$, respectively.
The eavesdropper receive antenna gain is set at $12\dBi$.
The thermal noise is set to $-174\dBmpHz$.
The residual \ac{SI} is fixed at $-110\dB$.
The \ac{DL} user, eavesdropper, and \ac{UL} user positions are set uniformly at $200\meters$, $175\meters$, and $150\meters$, respectively. 
The \ac{DFRC} \ac{BS} transmit power is $40\dBm$.

\begin{table}[!t]
\caption{Simulation Parameters\label{tab:table1}}
\centering
{
\begin{tabular}{|c||c|}
\hline
\textbf{Parameter} & \textbf{Value}\\
\hline
Number of cells & $1$ \\
\hline
Cell radius & $200\meters$ \\
\hline
Transmit power by \ac{DFRC} & $40\dBm$  \\
\hline
\ac{DFRC} transmit antenna gain & $25\dBi$  \\
\hline
\ac{UL} user antenna gain & $17\dBi$  \\
\hline
\ac{DFRC} receive antenna gain & $25\dBi$  \\
\hline
\ac{DL} receive antenna gain & $12\dBi$  \\
\hline
Eavesdropper receive antenna gain & $12\dBi$  \\
\hline
Carrier frequency & $28\GHz$  \\
\hline
Thermal noise& $-174\dBmpHz$ \\
\hline 
Shadowing & log-normal of $20\dB$ \\
\hline 
Pathloss exponent & $2$ (free space) \\
\hline 
Residual \ac{SI} & $-110\dB$ \\
\hline
$N_T$, $N_R$ & 12 \\
\hline
Antenna geometry & \ac{ULA} \\
\hline
Antenna spacing & $\frac{\lambda}{2}$ \\
\hline
Channel conditions & \makecell{Rayleigh \cite{9217488},\\ Rician ($K_r = 15\dB$) \cite{8207426}} \\
\hline
\ac{DL} user positions & uniformly at $200\meters$ \\
\hline
Eavesdropper positions & uniformly at $175\meters$ \\
\hline
\ac{UL} user positions & uniformly at $150\meters$ \\
\hline
\end{tabular}
}
\end{table}

\vspace{-0.5cm}
\subsection{Benchmark Schemes}
\label{sec:benchmark-schemes}
Throughout simulations, we compare our proposed method with the following three schemes: 
\begin{itemize}
	\item The isotropic design \cite{5524086}, denoted as \textit{ISO}, is a widely used benchmark to show the performance through isotropic power allocation.
In the \ac{FD} case, the power is uniformly distributed across the \ac{UL}, \ac{DL}, and \ac{AN} generation.
In particular, we set $\pmb{W} = \frac{P_{\tt{bud}}}{2} \frac{\mathcal{P}_{\pmb{\pmb{H}}_r}^\perp }{\Vert \mathcal{P}_{\pmb{\pmb{H}}_r}^\perp \Vert^2_F}$, where $\pmb{\pmb{H}}_r \in \mathbb{C}^{N_T \times L}$ contains all the \ac{DL} channel vectors $ \lbrace \pmb{h}_{r,\ell} \rbrace_{\ell = 1}^L$ stacked in the columns of $\pmb{\pmb{H}}_r$.
This design ensures that legitimate users will not be subject to any interference, however, eavesdroppers' reception may be harmed. 
On the other hand, the remaining available power is utilized for eigenbeamforming, i.e. $\pmb{V}_{:,\ell} =  \frac{\sqrt{P_{\tt{bud}}}}{\sqrt{2L}}\frac{\pmb{h}_{r,\ell}}{\Vert \pmb{h}_{r,\ell} \Vert}$.
	\item The no-\ac{AN} design, denoted as \textit{no-\ac{AN}}, is intended for communication-only tasks, where \ac{ZF} precoding is performed, namely $\pmb{V}  =  \pmb{H}_r (\pmb{H}_r^H \pmb{H}_r)^{-1}$ and $\pmb{W} = \pmb{0}$. 
Moreover, a uniform \ac{UL} power allocation is given to \ac{UL} users.
This benchmark is indicating when the main focus should be oriented towards communications-only tasks.
	\item The secure \ac{ISAC} design \cite{9199556}, denoted as \textit{SRCS}, represents an optimized \ac{ISAC} scheme where the optimization method aims at minimizing the \ac{SNR} towards an eavesdropper, under a communication \ac{SINR} constraint, a radar constraint and with a given power budget.
Note that this method is only applicable for half-duplex communication scenarios.

	\item \ac{SCA} without \ac{AN} is another interesting benchmark, where we run \textbf{Algorithm \ref{alg:alg1}} by setting $\pmb{W} = \pmb{0}$, and by forming a sub-problem of $(\mathcal{P}_{1.3}^{(m)})$ where $\pmb{W}$ is no longer treated as a variable simply because we set it to $\pmb{0}$. 
This benchmark is intended to show the importance of \ac{AN}.
\end{itemize}
The half-duplex scenario here is conducted with no \ac{UL} communication users, and is dedicated only for \ac{DL} secure communications, as well as sensing.
Therefore, only \ac{DL} \ac{ISAC} tasks are modeled.
This is a particular case of the \ac{FD} \ac{ISAC} model, i.e. by setting $K = 0$.
As the proposed method minimizes the power allocated for the secure \ac{FD} \ac{ISAC} system, we compute the total power consumed by the proposed method and use that power budget for all other benchmarks and settings, whether half-duplex or full-duplex, in order to have a fair comparison. Note that the power budget allocated to all the above three schemes is taken from the converged final output power of the proposed method. In other words, at each Monte Carlo trial, we first launch the proposed method, read the final power it had utilized, then pass that power budget to all the above three benchmarks to allow for a fair comparison. This is because the secrecy rate can be made arbitrarily large by increasing the transmission power. It is also worth noting that no-\ac{AN} design and the isotropic \ac{AN} are optimized for communications and secure communications, respectively. Therefore, no optimization is done regarding forming radar beams towards desired locations for further sensing processing at the \ac{DFRC} \ac{BS}.
	
\vspace{-0.3cm}
\subsection{Simulation Results}
\label{sec:simulation-results-sub-section}
\paragraph{Power Consumption}
In Fig. \ref{fig:pow_consumption_1}, we set $\sigma_n^2 = -80\dBm$, which is a typical \ac{mmWave} communication system thermal noise value \cite{8207426} corresponding to a noise figure of $7\dB$, and a \ac{PSD} noise floor of $-174\dBmpHz$.
We also set $\sigma_\ell^2 = -80\dBm$ and $\sigma_p^2 = -80\dBm$ as typical noise variances for \ac{mmWave} communication systems.
Furthermore, we fix $K = 2$ \ac{UL} communication users, $L = 1$ \ac{DL} communication user and $P = 1$ eavesdropper.
In this simulation, we set $\zeta_{p,\tt{DFRC}}^{\tt{Eve}} = 10^{-3}$, $ \zeta_{\ell}^{\tt{DL}} = 1, \forall \ell$ and $\zeta_k^{\tt{UL}} = 1$.
We run algorithm \textbf{Algorithm \ref{alg:alg1}} and average the obtained total consumed power per iteration over Monte Carlo trials. This simulation reveals how the total consumed power depends on the \ac{SINR} requirements, as well as the noise variance at the \ac{DFRC} \ac{BS}.
First of all, we highlight the power consumption as a function of secrecy requirements. More specifically, a more stringent constraint imposed on all the eavesdroppers has a direct impact on the total power consumption for a fixed noise power.
Indeed, for a target \ac{SINR} of 
$\zeta_{p,k}^{\tt{Eve}} = 1$, we observe that the total consumed power settles at about $-26.36 \dB$, 
whereas for a stricter \ac{SINR} requirement of $\zeta_{p,k}^{\tt{Eve}} = 0.032$, the total power converges to $-25.56 \dB$, which is an additional $0.8\dB$. 
Furthermore, setting $\zeta_{p,k}^{\tt{Eve}} = 0.001$, the total power converges to $-19.4 \dB$, which is an additional $6.16 \dB$.

Next, in Fig.\ref{fig:pow_consumption_2}, we increase the \ac{DFRC} \ac{BS} noise power to $\sigma_n^2 = -60\dBm$, which can be attained by either increasing \ac{UL} bandwidth, or at a high noise floor \cite{4266509}. We also observe a similar trend regarding the \ac{SINR} requirement as a function of total power consumed and convergence behaviour.
For example, 
at $\zeta_{p,k}^{\tt{Eve}} = 1$, the total consumed power stabilizes at $-6.43\dB$. 
Decreasing this \ac{SINR} requirement to $ 0.032$ increases the total consumed power by $0.92$, i.e. $-5.51\dB$. 
Finally, a very strict \ac{SINR} constraint on the eavesdroppers, $\zeta_{p,k}^{\tt{Eve}} = 0.001$, consumes $0.58\dB$, which is an additional $6.1 \dB$.
Besides the \ac{SINR} requirement, we also note that adding $20\dB$ of noise power is reflected within the total consumed power. For instance, setting $\zeta_{p,k}^{\tt{Eve}} = 0.001$ at $\sigma_n^2 = -80\dBm$ requires $20\dB$ less power than the same \ac{SINR} requirement at $\sigma_n^2 = -60\dBm$.
It is also worth noting that the algorithm converges in about $20$ iterations.
\paragraph{\ac{DL} Secrecy Behaviour}
In Fig. \ref{fig:DL_secrecy_1}, the \ac{DL} secrecy rate is analyzed as a function of $\sigma_\ell^2$.
We set a total of $P = 2$ eavesdroppers and $K = 9$ legitimate \ac{UL} users.
The number of antennas is fixed to $N_T = N_R = 12$.
We have fixed the \ac{SINR} requirements at $\zeta_\ell^{\tt{DL}} = \zeta_{k}^{\tt{UL}} = 10^4$ and $\zeta_{p,\tt{DFRC}}^{\tt{Eve}} = \zeta_{p,\tt{DFRC}}^{\tt{Eve}} = 10^{-5}$, for all $\ell,k,p$.
We set $\sigma_n^2 = -80\dBm$ and $\sigma_p^2 = -80 \dBm$.
It is observed that, regardless of channel conditions, the proposed method settles at about $13.28 \bpspHz$, which is inferred directly from the \ac{SINR} requirements, i.e. $\log_2(1 + 10^4) - \log_2(1+10^{-5})$. This reveals the consistency of our method regardless of the operating noise power level. 
The optimized SRCS method is somewhat stable as the \ac{DL} secrecy rate fluctuates around $11.9 \bpspHz$ for $L = 2$ for all channel conditions. As the number of \ac{DL} users increases, the secrecy rate decreases to around $10.98 \bpspHz$ for Rayleigh channels and $9.31 \bpspHz$ for Rician channel with $15\dB$ Rician factor \cite{8207426}.
This loss compared to the proposed method can be explained by the fact that SRCS is optimized to minimize the eavesdropper \ac{SNR} towards a given direction and under a power budget constraint.
Furthermore, the isotropic method is also stable as, for Rayleigh channels, it attains $3.66 \bpspHz$ for $L = 2$ and $0.98 \bpspHz$ for Rician channels with $K_{\tt{r}} = 15\dB$. The secrecy rate decreases with increasing $L$ as a drop of $2.51\bpspHz$ is observed for $L = 2$ in case of Rayleigh channels and a drop of $0.84\bpspHz$ in case of Rician channels. 
Moreover, the no-\ac{AN} method depicts a different behavior. More specifically, in cases of high \ac{DL} \ac{SNR}, i.e. low $\sigma_\ell^2$, the no-\ac{AN} is generally better than the isotropic allocation, as for Rayleigh channels, a $5.2\bpspHz$ secrecy is attained for $L=2$, and $3.2 \bpspHz$ for $L=5$, whereas secrecy rates of $1.72\bpspHz$ ($L=2$) and $1.63\bpspHz$ ($L = 5$) is achieved for Rician channels.
In the low \ac{SNR} regime, $\sigma_\ell^2 > -55 \dBm$, we see that the \ac{DL} secrecy rate for no \ac{AN} decays to zero, as expected. This is because the power is not optimally distributed, and hence the \ac{SINR} at the eavesdroppers will exceed that of the legitimate users after a certain $\sigma_\ell^2$.
In addition, we highlight two main insights: 
First, the proposed method performs better than the isotropic and the no-\ac{AN} methods due to the fact power and beamfomers are optimized to achieve target secrecy rates under minimal power.
Second, we interestingly observe that strong \ac{LoS} channels may be harmful for the secrecy rates, as higher channel correlations are achieved, which can deteriorate the overall secrecy performance.

In Fig. \ref{fig:DL_secrecy_2}, we study the \ac{DL} secrecy rate performance as a function of noise power at the eavesdroppers.
We set the same simulation parameters as those in Fig. \ref{fig:DL_secrecy_1}.
A similar trend is noticed for all the methods, except for the no \ac{AN} case.
In particular, when the noise at the eavesdroppers exceeds $-30\dBm$, the \ac{DL} secrecy rate for the no \ac{AN} case increases and can outperform the \ac{DL} secrecy rate of the proposed method.
This can be explained by the fact that the scheme without \ac{AN} utilizes all the available power to optimize the communication performance, without having to worry about the eavesdroppers, which are already experiencing high noise power levels. Therefore, in such a noisy regime, it may be enough to focus on maximizing the \ac{DL} \ac{SINR} without allocating power towards \ac{AN} beamforming.
Another point to highlight is that the proposed \ac{SCA} is optimized to target a minimum acceptable \ac{DL} \ac{SINR} regardless of the noise level $\sigma_p^2$, therefore, in order to achieve a higher secrecy rate, we would have to increase $\rho^{\tt{DL}}$, similar to the simulations corresponding to Fig. \ref{fig:UL_secrecy_1}.
SRCS, as well as the proposed method, seek to attain the required \ac{SINR} values as part of the optimization procedure, whereas the no \ac{AN} performs \ac{ZF} precoding and can attain better performance only when the eavesdroppers operate at a high noise power level.

In Fig. \ref{fig:DL_secrecy_3}, our goal is to study the influence of \ac{UL} \ac{SINR} requirements on the \ac{DL} secrecy rate. We fix the same simulation parameters as in Fig. \ref{fig:DL_secrecy_2}.
This simulation shows how the \ac{UL} \ac{SINR} requirement impacts the \ac{DL} secrecy rate. 
The SRCS method, being half-duplex, will profit from all the power spent to improve its \ac{DL} secrecy rate. But, as observed, the SRCS method settles around $11.96\bpspHz$ for $L=2$ and $11.07\bpspHz$ for $L =5$, whereas the proposed method supplies a stable \ac{DL} secrecy rate of $13.28 \bpspHz$, for any given \ac{UL} \ac{SINR} requirement.
It is also observed that with increasing number of \ac{DL} users, all but the proposed method deteriorate in terms of secrecy rate.
The \ac{DL} secrecy rate achieved by the no-\ac{AN} design method increases with increasing $\rho^{\tt{UL}}$ since more power is being allocated to the system. However, the \ac{DL} secrecy rate attained by no \ac{AN} scheme depends on the number of legitimate \ac{UL} and \ac{DL} communication users. Generally speaking, the lower $L$ is, the higher the \ac{DL} secrecy rate that can be achieved.

\paragraph{\ac{UL} Secrecy behaviour}
In Fig. \ref{fig:UL_secrecy_1}, we study the \ac{UL} secrecy rate behavior versus the \ac{SINR} requirement on the \ac{UL}, i.e. $\rho^{\tt{UL}}$.
We distinguish two distance/position distributions, where the \ac{UL} users are placed on a circle of radius $150\meters$ for the first distribution, whereas the same \ac{UL} users are placed on a straight line uniformly distributed on $[1,500] \meters$.
In the first distribution, the performance of all three methods perform alike, as a uniform power allocation onto the \ac{UL} users suffices to achieve a desired \ac{SINR} on the \ac{UL} users.
In the second distribution, however, for the same total consumed power, the proposed \ac{SCA} method can achieve a better \ac{UL} secrecy rate, i.e. an additional $1.38 \bpspHz$, as the no-\ac{AN} and isotropic schemes allocate equal amounts of power to all the \ac{UL} users.  
Note that the SRCS method is not present because it is dedicated for half-duplex \ac{ISAC} communication systems. 
\paragraph{\ac{ISMR} impact on Beampattern Design} 
In Fig. \ref{fig:bp_1}, we study the resulting normalized beampattern after executing \textbf{Algorithm \ref{alg:alg1}} for different \ac{ISMR} values.
The normalized beampattern is defined as 
\vspace{-0.3cm}
\begin{equation}
P(\theta) 
	=
	\frac{\pmb{a}^H(\theta)
	\widehat{\pmb{R}}_{\pmb{x}\pmb{x}}
	\pmb{a}(\theta)}{\max_{\theta} \pmb{a}^H(\theta)
	\widehat{\pmb{R}}_{\pmb{x}\pmb{x}}
	\pmb{a}(\theta)},
\end{equation}
where $\widehat{\pmb{R}}_{\pmb{x}\pmb{x}} = \widehat{\pmb{V}}\widehat{\pmb{V}}^H  + \widehat{\pmb{W}}$.
We evaluate $P(\theta)$ for different values of $\theta$ in order to obtain the beampattern. 
For the \ac{SINR} requirements on the legitimate users, we set $\zeta_{k}^{\tt{UL}}  = \zeta_{\ell}^{\tt{DL}} = 10^2$.
As for the \ac{DL}/\ac{UL} \ac{SINR} on evesdroppers, we fix $\zeta_{p,k}^{\tt{Eve}} = \zeta_{p,\tt{DFRC}}^{\tt{Eve}} = 0.1$.
We also set all noise powers to $-80\dBm$.
This simulation shows how the $\ISMR_{\tt{max}}$ can help control the mainlobe to sidelobe ratio power.
Indeed, when setting $\ISMR_{\tt{max}} = -30\dB$, we observe that the sidelobe power is roughly $10\dB$ more than the beampattern corresponding to that of $\ISMR_{\tt{max}} = -20\dB$ and $10\dB$ less than that of $\ISMR_{\tt{max}} = -40\dB$.
It is worth noting that a less strict \ac{ISMR} constraint, allows for a higher achievable secrecy, which will be clarified in the next simulation.
\paragraph{Sensing-Secrecy \ac{ISAC} Trade-off} 
In Fig. \ref{fig:tradeoff}, we study the sensing-secrecy trade-off, where the achievable \ac{ISMR} inverse is plotted against the secrecy rate.
 All noise powers are set to a level of $-80\dBm$, and the number of antennas is set to $N_T = N_R = 12$.
 It is revealed that the secrecy rate improves on the expense of increasing the \ac{ISMR}. 
Studying the trade-off at extreme cases for $P = 3$, we observe that an $\ISMR^{-1}$ can only achieve a secrecy rate of $0.74 \bpspHz$, which is the case of pointy beams where the echo return experiences a power boost good clutter rejection properties, whereas relaxing the $\ISMR^{-1}$ anywhere below $-20\dB$ can achieve a secrecy rate of $2.76 \bpspHz$.
Increasing the number of eavesdroppers impacts the tradeoffs, as lower secrecy rates are achieved via the same resources. For instance, an additional evesdropper lowers the secrecy rate to $2.35 \bpspHz$ when $\ISMR^{-1} \leq -20\dB$.
We conclude that the $\ISMR_{\tt{max}}$ can also allow us to achieve favorable tradeoffs to attain a desired secrecy rate, while performing radar tasks.
\paragraph{Importance of \ac{AN}} In Fig. \ref{fig:tradeoff2}, we study the performance of the proposed method with and without \ac{AN}.
In the first subplot of Fig. \ref{fig:tradeoff2}, we study the secrecy rate as a function of total consumed power.
We can observe a significant gain in terms of total consumed power for nearly all secrecy rates. 
For example, a target secrecy rate of $3.79 \bpspHz$ requires $8.35 \dB$ without using \ac{AN}, whereas the same secrecy rate consumes only $-1.92 \dB$ for the proposed method with \ac{AN}, which translates to a total power gain of about $10.27\dB$.
This can be explained through the additional subspace created by the \ac{AN}, which can indeed save notable amounts of power to achieve a desired secrecy rate.

In the second subplot of Fig. \ref{fig:tradeoff2}, we study the feasibility rate of both schemes, which is defined as \cite{10018908}

\begin{equation}
	\feasibilityrate
	=
	\frac{\# \text{ of feasible solutions}}{\# \text{ of cases}}.
\end{equation}
It is observed that for secrecy rate below $2.73\bpspHz$, the feasibility rate of both schemes is nearly $1$, i.e. we can always find beamformers and power values (as well as \ac{AN} covariance matrix for the scheme with \ac{AN}) to target the desired secrecy rates. 
When the secrecy rate goes beyond $2.73\bpspHz$, we can see that the proposed scheme with \ac{AN} outperforms that without \ac{AN}.
For instance, for a fixed desired feasibility rate of $0.93$, i.e. a feasible solution can be found $93\%$ of the time, the proposed scheme without \ac{AN} can only achieve a secrecy rate up to $3.79\bpspHz$, whereas the proposed scheme with \ac{AN} can nearly double that secrecy rate to go up to $7.35\bpspHz$.
This is due to the additional degrees of freedom introduced by the \ac{AN}, which enables us to achieve much higher secrecy rates for more cases.

\vspace{-0.2cm}
\section{Conclusions and Research Directions}
\label{sec:conclusions}
We have provided a \ac{FD} \ac{ISAC} system model, where a unified waveform for sensing and secure communications is transmitted in the \ac{DL} to maintain desired \ac{UL} and \ac{DL} secrecy rate performance, whilst guaranteeing desired \ac{ISMR} radar levels. Under the proposed model, we proposed a power efficient \ac{ISAC} optimization framework to jointly satisfy \ac{UL}/\ac{DL} secrecy rates, under a maximal acceptable \ac{ISMR}, which further aids the \ac{DFRC} \ac{BS} at probing sensing information from the eavesdroppers. The total power minimization problem is intended to economize the overall transmitted \ac{UL}/\ac{DL} power and the artificial noise generation power. Due to the intractability introduced by the non-convex problem, we propose iterating via successive convex approximations, which is deemed suitable due to the low number of iterations spent by the method, and its capability in simultaneously guaranteeing \ac{UL}/\ac{DL} secrecy rates and \ac{ISMR} levels. Our results support the effectiveness of the method, as compared to classical secure beamforming approaches and secure \ac{ISAC} methods. 
Future work include extensions towards an energy-efficient optimization problem that can also be formulated under the proposed \ac{FD} \ac{ISAC} system model. Indeed, energy efficiency emanated as a crucial key performance indicator for the deployment of sustainable and green wireless networks \cite{7446253}. 
Furthermore, an extension to the imperfect \ac{CSI} case will also be part of our future work.

In addition to the security vulnerabilities in future wireless communications mentioned in this paper, the survey in \cite{10024766} highlights a number of critical problems over \ac{B5G} wireless networks and \ac{IoT}. 
For instance, edge networks face an onslaught of risks and attacks because of their limited processing power, memory capacity, battery life, and network bandwidth, in addition to threats encountered in \ac{IoT}, such as impersonation attacks and routing attacks. 
As highlighted by \cite{10024766}, and references therein, \ac{DL}-based techniques seem to be a promising approach for future wireless communications, e.g., physical-layer authentication \cite{9517121}.
In some cases, however, learning can be effective for online usage due to its low online complexity, but it can suffer from high computational complexity when training, whereas a fully distributed training approach can suffer from convergence issues due to the partial availability of state information for training. 
In contrast to a centralized and insecure training approach for wireless communications, \ac{EL} has been proposed to distribute the implementation of \ac{ML} algorithms over multiple edge devices, thereby reducing computational complexity and malicious attacks. 
Furthermore, the use of \ac{EL} methods in a range of future communication systems, including \ac{B5G} wireless networks, is also being considered.
For instance, \cite{zhu2019deep} aims at communications and privacy protection via \ac{FL}, whereas \cite{8896930} detects network intrusion.
As future work, we can leverage the philosophy of \ac{EL} in order to distribute the computations of the proposed \ac{SCA}-based algorithm for minimal power transmissions in secure \ac{FD} \ac{ISAC}, which can contribute to lower computations due to the relatively large number of required iterations.

\vspace{-0.5cm}
\appendices
\label{sec:appendix}
\section{Rank-1 Optimality}
\label{appendix:proof-rank-1}
We start by expressing the Lagrangian function of the optimization problem given at iteration $m$ in \eqref{eq:problem1_3}. It can be expressed as
\begin{equation*}
	\begin{split}
		&\mathcal{L}(\pmb{V}_1 \ldots \pmb{V}_L ,\pmb{W},\pmb{p} , \pmb{a},\pmb{b}, \pmb{c},\pmb{d},\zeta,\pmb{\Pi}_0 \ldots \pmb{\Pi}_L,\pmb{\pi}) \\ & = -\sum\limits_{\ell = 1}^L\trace(\pmb{\pmb{V}}_\ell) - \trace(\pmb{W}) - \pmb{1}^T \pmb{p} 
		 + 
	\trace(\pmb{\Pi}_0 \pmb{W} )
	+
	\sum\limits_{\ell=1}^L
	\trace(\pmb{\Pi}_\ell \pmb{V}_\ell )
		\\ &
		+ \pmb{\pi}^T \pmb{p}
		+ \sum\nolimits_{\ell = 1}^L a_\ell\phi_{1,\ell}
		+ \sum\nolimits_{k = 1}^K b_k\phi_{2,k}
		+ \sum\nolimits_{p = 1}^P c_p\phi_{3,p} \\ 
		&+ \sum\nolimits_{p = 1}^P\sum\nolimits_{k=1}^K d_{p,k}\phi_{4,p,k} +  \zeta\phi_{5},
	 \end{split}
\end{equation*}
\noindent where the  Lagrangian dual variables satisfy $a_\ell \geq 0, b_k \geq 0, c_p \geq 0, d_{p,k} \geq 0 , \zeta \geq 0, \pmb{\Pi}_\ell \succeq \pmb{0}$ for all $\ell,k,p$. The functions $\phi_{1,\ell},\phi_{2,k},\phi_{3,p},\phi_{4,p,k},\phi_5$ are obtained by mapping the constraints in \eqref{eq:problem1_3} under the form $\phi \geq 0$. Due to compact representation, we have omitted the dependence of $\phi$ on the optimization variables and the iteration number $m$. To this extent, we can now define the Lagrange dual function is then defined as follows
\vspace{-0.05cm}
\begin{equation}
	\begin{split}
& g(\pmb{a},\pmb{b}, \pmb{c},\pmb{d},\zeta,\pmb{\Pi}_0 \ldots \pmb{\Pi}_L,\pmb{\pi}) 
\\ &= \max\limits_{ \mathcal{S} } \mathcal{L}(\pmb{V}_1 \ldots \pmb{V}_L ,\pmb{W},\pmb{p} , \pmb{a},\pmb{b}, \pmb{c},\pmb{d},\zeta,\pmb{\Pi}_0 \ldots \pmb{\Pi}_L,\pmb{\pi}), 
	\end{split}
\end{equation}
where the set $\mathcal{S}$ is defined as $\mathcal{S} = \lbrace \pmb{V}_\ell \succeq \pmb{0},  \pmb{W} \succeq \pmb{0},  \pmb{p} \geq \pmb{0} \rbrace $.
But since problem $(\mathcal{P}_{1.3}^{(m)})$ is a convex optimization problem with zero duality gap, we can solve the problem via its dual
\vspace{-0.1cm}
\begin{equation}
(\mathcal{D}):
	\min\limits_{ \bar{\mathcal{S}} } g(\pmb{a},\pmb{b}, \pmb{c},\pmb{d},\zeta,\pmb{\Pi}_0 \ldots \pmb{\Pi}_L,\pmb{\pi}),
\end{equation}
\noindent where $\bar{\mathcal{S}} = \lbrace \zeta \geq 0, \pmb{a},\pmb{b},\pmb{c},\pmb{d}, \pmb{\pi}  \succeq 0, \pmb{\Pi}_\ell \succeq 0,  \forall \ell \rbrace$.
Let the optimal solution of problem $(\mathcal{D})$ as $\pmb{a}^{\opt},\pmb{b}^{\opt}, \pmb{c}^{\opt},\pmb{d}^{\opt},\zeta^{\opt},\pmb{\Pi}_0^{\opt} \ldots \pmb{\Pi}_L^{\opt},\pmb{\pi}^{\opt}$. This means that the optimal $\ell^{th}$ \ac{DL} beamforming matrix, the \ac{AN} covariance matrix, and \ac{UL} power allocation denoted as $\pmb{V}_\ell^{\opt},\pmb{W}^{\opt},\pmb{p}^{\opt}$ also maximizes $\mathcal{L}(\pmb{V}_1 \ldots \pmb{V}_L ,\pmb{W},\pmb{p} ,\pmb{a}^{\opt} \ldots \pmb{\pi}^{\opt}) $, which means that we can find the optimal solution $\pmb{V}_\ell^{\opt}$ by solving the following optimization problem
\vspace{-0.05cm}
\begin{equation}
\label{eq:dual-formulation}
\begin{split}
	\max\limits_{  \pmb{V}_\ell \succeq \pmb{0}  }
	\Bigg\lbrace
	&-\trace(\pmb{V}_\ell) 
	+ \frac{a_\ell}{\zeta_{\ell}^{\tt{DL}}}\trace(\pmb{V}_\ell\pmb{H}_{r,\ell})+
	\trace(\pmb{\Pi}_\ell \pmb{V}_\ell)
	\\ &
	- \sum\nolimits_{k = 1}^K b_k
	  \pmb{h}^H_{t,k}
	[\pmb{\Phi}_k^{(m)}]^{-1}
	\pmb{C}\pmb{V}_\ell\pmb{C}^H
	[\pmb{\Phi}_k^{(m)}]^{-1}
	\pmb{h}_{t,k} 
	\\ &
	- \sum\limits_{p=1}^P 
	\big(  \frac{c_p \vert \alpha_p \vert^2}{\zeta_{p,\tt{DFRC}}^{\tt{Eve}} }  
	 -
	\sum\limits_{k=1}^K d_{p,k}
	\vert \alpha_p \vert^2
	 \big)
	 \pmb{a}_{N_T}^H(\theta_p)  \pmb{V}_\ell
	 \pmb{a}_{N_T}(\theta_p)
	\\ &
	+
	\zeta\Big( 
	\ISMR_{\tt{max}} \times \trace\Big( \pmb{V}_\ell \pmb{A}_m \Big)
	-
	\trace\Big( \pmb{V}_\ell \pmb{A}_s \Big)
	\Big) 
	  \Bigg\rbrace,
\end{split}
\end{equation}
where terms that are independent of $\pmb{V}_\ell$ are ignored. We can compactly express \eqref{eq:dual-formulation} as follows
\begin{equation}
\label{eq:dual-maximization}
	\max\limits_{  \pmb{V}_\ell \succeq \pmb{0}  }
	\trace(\Phi \pmb{V}_\ell ) 
	-
	\trace \Big\lbrace (\pmb{I} + \pmb{\Gamma} + \zeta^{\opt} \pmb{A}_s) \pmb{V}_\ell  \Big\rbrace,
\end{equation}
where
\vspace{-0.05cm}
\begin{align}
	\Phi &= \frac{a_\ell^{\opt}}{\zeta_{\ell}^{\tt{DL}}}\pmb{H}_{r,\ell}
	 + \pmb{\Upsilon} + \zeta^{\opt}\ISMR_{\tt{max}} \pmb{A}_m + \pmb{\Pi}_\ell^{\opt},\\
	 \pmb{\Upsilon} &= \sum\nolimits_{p=1}^P
	\sum\nolimits_{k=1}^K
	d_{p,k}^{\opt}
	\vert \alpha_p \vert^2 \pmb{a}_{N_T}(\theta_p)\pmb{a}_{N_T}^H(\theta_p),\\
	\label{eq:ammended-gamma-expression}
	\begin{split}
		\pmb{\Gamma} &
	=
	\sum\nolimits_{k = 1}^K b_k^{\opt}
	\pmb{C}^H
	[\pmb{\Phi}_k^{(m)}]^{-1}
	\pmb{h}_{t,k}
	  \pmb{h}^H_{t,k}
	[\pmb{\Phi}_k^{(m)}]^{-1}
	\pmb{C}  \\&+ \sum\nolimits_{p=1}^P 
	 \frac{c_p^{\opt} \vert \alpha_p \vert^2}{\zeta_{p,\tt{DFRC}}^{\tt{Eve}} }
	 \pmb{a}_{N_T}(\theta_p)\pmb{a}_{N_T}^H(\theta_p).
	\end{split} 
\end{align}	 
Note that \eqref{eq:dual-maximization} is bounded above because both $\Phi$ and $\pmb{I} + \pmb{\Gamma} + \zeta^{\opt} \pmb{A}_s$ are positive semi-definite. 
Now, since $\pmb{I} + \pmb{\Gamma} + \zeta^{\opt} \pmb{A}_s \succeq \pmb{0}$, there is a unique Hermitian positive definite matrix $\pmb{R}_k^{\frac{1}{2}}$, such that $ \pmb{I} + \pmb{\Gamma} + \zeta^{\opt} \pmb{A}_s = \pmb{R}^{\frac{1}{2}}\pmb{R}^{\frac{1}{2}}$ \cite{horn2012matrix}. Using square root matrix decomposition, we can write our cost in \eqref{eq:dual-maximization} as 
\begin{equation}
	\label{eq:cost-with-sqrt-decom}
	 \max\limits_{  \widetilde{\pmb{V}}_\ell \succeq \pmb{0}  }
	 \trace \Big\lbrace
	  \pmb{R}^{-\frac{1}{2}} \Phi \pmb{R}^{-\frac{1}{2}} \widetilde{\pmb{V}}_\ell 
	  \Big\rbrace
	    -
	 \trace \Big\lbrace \widetilde{\pmb{V}}_\ell \Big\rbrace,
\end{equation} 
where $ \widetilde{\pmb{V}}_\ell = \pmb{R}^{\frac{1}{2}}\pmb{V}_\ell\pmb{R}
^{\frac{1}{2}}$. Now, suppose that the optimal solution, denoted hereby as $\widetilde{\pmb{V}}_\ell^{\opt}$, is not $\rank$-one, i.e. $\rank$ $S > 1$. Then, employing the eigenvalue decomposition, we have that $\widetilde{\pmb{V}}_\ell^{\opt} = \sum\nolimits_{n=1}^S \beta_n \widetilde{\pmb{v}}_n \widetilde{\pmb{v}}_n^H$ where $\beta_n \geq 0$ and $\Vert \widetilde{\pmb{v}}_n \Vert = 1$, the cost in \eqref{eq:cost-with-sqrt-decom} can be written as
\begin{equation}
\label{eq:cost-with-sqrt-decom-2}
\max\limits_{ \Vert \widetilde{\pmb{v}}_n \Vert = 1  , \beta_n \geq 0 }
	\sum\nolimits_{n=1}^S
	\beta_n \big( \Vert \Phi^{\frac{1}{2}} \pmb{R}^{-\frac{1}{2}}\widetilde{\pmb{v}}_n \Vert^2 - 1 \big).
\end{equation}
Notice that the cost in \eqref{eq:cost-with-sqrt-decom-2} is upper-bounded as follows
\begin{equation*}
\begin{split}
&	\sum\limits_{n=1}^S
	\beta_n \big( \Vert \Phi^{\frac{1}{2}} \pmb{R}^{-\frac{1}{2}}\widetilde{\pmb{v}}_n \Vert^2 - 1 \big)
	 \leq  
	\sum\limits_{n=1}^S 
	\beta_n \big( \Vert \Phi^{\frac{1}{2}} \pmb{R}^{-\frac{1}{2}}\widetilde{\pmb{v}}_{\hat{n}} \Vert^2 - 1 \big),
\end{split}
\end{equation*} 
where $\Phi^{\frac{1}{2}}$ is the square root decomposition of $\Phi$. Indeed, we have upper bounded all terms of the form $ \Vert \Phi^{\frac{1}{2}} \pmb{R}^{-\frac{1}{2}}\widetilde{\pmb{v}}_n \Vert^2$ by $ \Vert \Phi^{\frac{1}{2}} \pmb{R}^{-\frac{1}{2}}\widetilde{\pmb{v}}_{\hat{n}} \Vert^2$ where $\hat{n} = \argmax\nolimits_{n = 1 \ldots S} \Vert \Phi^{\frac{1}{2}} \pmb{R}^{-\frac{1}{2}}\widetilde{\pmb{v}}_n \Vert^2 $. But, this bound is achieved by the following $\rank$-one form $\widetilde{\pmb{V}}_\ell^{\rank-1} = (\sum\nolimits_{n=1}^S \beta_n ) \widetilde{\pmb{v}}_{\hat{n}} \widetilde{\pmb{v}}_{\hat{n}}^H$,
which is a contradiction of the assumption that $S > 1$. Therefore, $\widetilde{\pmb{V}}_\ell^{\opt} $ has to be $\rank$-1 optimal. Now, since we have ${\pmb{V}}_\ell^{\opt} = \pmb{R}^{-\frac{1}{2}}\widetilde{\pmb{V}}_\ell^{\opt}\pmb{R}^{-\frac{1}{2}}$, we must also have that ${\pmb{V}}_\ell^{\opt} $ is $\rank$-one optimal according to $\rank(\pmb{A}\pmb{B}) \leq \min\lbrace \rank(\pmb{A}),\rank(\pmb{B})\rbrace$, which finalizes the proof.

\section*{Acknowledgment}
This work has been supported by Tamkeen and the Center for Cybersecurity under the NYU Abu  Dhabi Research Institute Award G1104.

\bibliographystyle{IEEEtran}
\bibliography{refs}

\vfill

\end{document}